\documentclass[a4paper,11pt]{article}
\pdfoutput=1 

\usepackage{jcappub} 
\usepackage{float}
\usepackage[T1]{fontenc} 

\title{WIMP decay as a possible Warm Dark Matter model}

\author[a,1]{Abineet Parichha,\note{Corresponding author.}}
\author[b]{Shiv Sethi}

\affiliation[a]{Indian Institute of Science Education and Research, Mohali, India}
\affiliation[b]{Raman Research Institute, Bangalore, India}

\emailAdd{ms18094@iisermohali.ac.in}
\emailAdd{sethi@rri.res.in}

\abstract{The  Weakly Interacting Massive Particles (WIMPs) have long been the favoured  Cold Dark Matter (CDM) candidate in the standard  $\Lambda$CDM model. However, owing to  great improvement in the experimental sensitivity in the past  decade, some parameter space of the Supersymmetric (SUSY)-based WIMP model is ruled out. In addition, a massive stable WIMP as the CDM particle is also at variance with other astrophysical observables at small scales. We consider a model that addresses both these issues. In the model, the WIMP decays into a massive particle and radiation. We study the background evolution and the first order perturbation theory (coupled Einstein-Boltzmann equations) for this model and show that the dynamics can be captured by a single parameter $r=m_L/q$, which is the ratio of the lighter mass and the 
comoving momentum of the decay particle. We incorporate the relevant equations in the existing  Boltzmann code CLASS to compute the matter power spectra and  Cosmic Microwave Background (CMB) angular power spectra. The decaying  WIMP model is akin to a non-thermal Warm Dark Matter (WDM) model and suppresses matter power at small scales, which could alleviate several issues that plague the CDM model at small scales. We compare the predictions
of the model with CMB and galaxy clustering data. As the model deviates from the $\Lambda$CDM model at small scales, the evolution of the collapse fraction of matter in the universe is compared with the high-redshift  Sloan Digital Sky Survey (SDSS) HI data. Both these data sets   yield $r \gtrsim 10^6$, which
can be translated into the bounds on other parameters. In particular, we obtain the following lower bounds on the thermally-averaged self-annihilation cross-section of WIMPs, $\langle\sigma v \rangle$, and the lighter mass: $\langle \sigma v  \rangle \gtrsim 4.9\times 10^{-34} \, \rm cm^3 \, sec^{-1}$ and $m_L \gtrsim 2.4 \, \rm keV$. 
The lower limit on $m_L$ is comparable to constraints  on  the mass of thermally-produced WDM  particle. The  limit on the self-annihilation cross-section greatly expands the available parameter space as compared to the stable WIMP scenario.
}

\keywords{Dark matter theory, cosmological perturbation theory, particle physics-cosmology connection, cosmological parameters from LSS}

\begin{document}
\maketitle
\flushbottom

\section{Introduction}
\label{sec:intro}

The standard cosmological model has proved to be very  successful
during  the past two decades. Among other probes, the  measurement of CMB temperature and polarization
anisotropies,  galaxy clustering as revealed by large galaxy surveys, and the detection of high-redshift supernova~1a  have been
key to this success \citep{hinshaw2013nine,aghanim2020,abolfathi2018fourteenth,conley2010supernova}. An important ingredient of the concordance $\Lambda$CDM  model is the cold dark matter. Its properties are
indirectly inferred based on many  observations covering a wide range of 
 length scales, from sub-galactic to cosmological,  and epochs of the universe 
(e.g. \cite{Zwicky,Rubin,aghanim2020,abolfathi2018fourteenth,Bartelmann:1999yn}).

While CMB and galaxy clustering observations show that the  CDM is a good
candidate of dark matter for scales $k < 0.1 \, \rm Mpc^{-1}$,  there exist  long-standing astrophysical  issue with the model at smaller scales. The CDM N-body simulations  predict  an order of magnitude larger number of satellite galaxy of
the Milky way as compared to the observed number\citep{moore1999dark,small_scale_crisis}. The  N-body simulations based on the $\Lambda$CDM model  predict
a cuspy profile at the center of galaxies but the observed profile is flat \citep{Cuspy_halo}.   Another issue  to emerge from the comparison of CDM N-body  simulations  with observations is the  ``too big to fail'' problem 
\cite{Garrison-Kimmel:2014vqa,BoylanKolchin:2011de}.  All these issues provide  motivation  to consider alternatives,  which  reproduce the successes  of the CDM model on   cosmological scales but differ from the $\Lambda$CDM  model at small scales. The discovery of many  high-redshift galaxies with unusually high stellar mass by  JWST  (e.g. \cite{2023MNRAS.518.6011D,2023Natur.616..266L}) could be pointing at models other than the  $\Lambda$CDM model. However, it is equally likely that the observed behaviour is owing to much higher star formation efficiency at high redshifts (e.g. \cite{2023MNRAS.518.6011D} and references therein). This issue is still being debated so we 
shall not attempt to study the possible implication of these results in this paper.

In the $\Lambda$CDM model, one of the possible candidates for the CDM particle is  the Weakly Interacting Massive Particle (WIMP).   Such  massive, stable, particles arise naturally in the   supersymmetric extension of the standard model of particle physics.   The theory  predicts the     CDM energy density  infrerred from cosmological observables for  self-annihilation cross-section $\langle \sigma v\rangle \simeq 3 \times 10^{-26}\rm{cm^3s^{-1}}$ and   WIMP masses in the range 10--1000 GeV (WIMP miracle, e.g.\cite{Craig:2015xla}). This coincidence has spurred many direct 
\cite{PhysRevLett.121.111302,Ahmed:2010wy,PhysRevLett.118.251302}, indirect\cite{Adriani:2010rc,FermiLAT:2011ab,Aguilar:2007yf} and 
collider \cite{Goodman:2010yf,Fox:2011pm} searches  of the WIMP. 
However, even after extensive laboratory and astronomical searches, the WIMP  dark matter is yet to be detected. 
Direct lab searches, based on the scattering of dark matter particles with heavey nuclei,  are sensitive to both
spin-dependent and spin-independent interactions (e.g. \cite{PhysRevLett.121.111302,PhysRevLett.118.251302}). These experiments have achieved unprecedented precision in the recent years and have 
begun to rule out parameter-space favoured by supersymmetric extensions of  the standard model (e.g. \cite{2012EPJC...72.2020B,2018RPPh...81f6201R,2019JPhG...46j3003S,2014JHEP...09..081S}). In light of this fact, one needs
to explore extensions to WIMPs-based models. In this paper, we consider such  a model in which the WIMP decays and one of the decay products of WIMPs acts as cold dark matter at late times. From theoretical perspective, this scenario allows us to expand the permissible space of parameters. Additionally, we show that such a model leave observable signatures on CMB and galaxy clustering data and has a bearing on the small-scale issue with the CDM  paradigm and therefore is potentially detectable by the current and future  cosmological data.

In section \ref{sec:wdm}, we discuss our model in detail. We derive the phase space distribution function of the decay products of WIMPs and  show how different phases of this process impact the background evolution of the universe---
from the production of WIMPs after freeze out in the very early universe to the evolution of the decay products of WIMPs. In section \ref{sec:pert_theory}, we derive the first order perturbation theory for the decay products in Newtonian-conformal gauge and discuss
the novel features of this scenario. We also outline the steps followed to incorporate our model in the existing code CLASS\cite{CLASS_2011}. 
In section~\ref{sec:resall} our  main  results are presented. In section \ref{sec:data_analysis}, we consider a number of data sets to test our model:  Planck CMB data,  SDSS (BOSS) data,  and the evolution of neutral hydrogen mass density from damped Lyman-$\alpha$ data. We use likelihood data of the  two-point angular correlation functions of temperature, polarisation, and lensing potential fluctuations from Planck 2018 CMB  data \cite{aghanim2020} and baryon acoustic oscillation (BAO) data of luminous red galaxy (LRG) distribution from BOSS \cite{BOSS_bao}. We  determine  posterior probabilities on relevant parameters using Montepython \cite{MontepyOriginal_2013,MontePy3_2018} MCMC codes. The CMB and BAO data allow us 
to compare our model with the  data for scales $k < 0.2 \, \rm Mpc^{-1}$.   To test at smaller scales, we  compute the evolution of collapsed  fraction of matter in the universe and compare our theoretical prediction against  the  inferred collapsed fraction of neutral hydrogen obtained from the  SDSS Lyman$\alpha$ data \cite{SDSS_dr7, Peroux_HI}.  Section \ref{sec:conclusion} is reserved for summarizing our main results and  concluding remarks. Unless specified otherwise, we consider the spatially-flat cosmological model with Planck best-fit  cosmological parameters 
(\cite{aghanim2020}).

\section{Decaying WIMP}
\label{sec:wdm}

We propose a scenario in which   non-relativistic  WIMPs of mass $m_H$ decay  into a lighter  particle of mass $m_L$ and a massless particle: $m_H \rightarrow m_L + \text{radiation}$. For simplicity, we assume the decay to be instantaneous at $a = a_d$. The decay is assumed to be triggered after the freeze-out. This allows us to treat the processes of the  freeze-out and 
the decay separately. We discuss the potential impact of this assumption in section~\ref{sec:concl}.

Assuming the WIMPs to be highly non-relativistic at the time of decay (or the ratio of their momentum to energy is negligible), the  comoving momentum of the lighter particle (we refer to the lighter particle as warm dark matter (WDM) in the rest of the paper as our model is akin to non-thermally produced warm dark matter particles) and the decay radiation is given by: $q_L = q_{\rm dr} = a_d (m_H^2 - m_L^2) / 2m_H$. The abundance of the WDM and the radiation particles is equal to the relic abundance $N$ of WIMPs. Thus,  from relativistic kinematics, we can construct the following phase-space distribution function for the   decay products---WDM and radiation---of WIMPs:
\begin{align}
    \label{eq:psd_WDM}f_0(q) &= \frac{N}{4\pi q^2}\delta\Big(q-\frac{a_d(m_H^2 - m_L^2)}{2m_H}\Big)
\end{align}
 We note that the WDM particles are not in thermal equilibrium. WDM particles can be either relativistic or  non-relativistic at the epoch of decay. For  $m_L \ll m_H$, the WDM particle is relativistic but it could 
be non-relativistic if $m_L \simeq m_H$.  $N$ denotes the comoving number density of 
WDM particles, defined such that: $\int f_0(q) d^3q = N$.  In this paper, we follow the momentum and energy coordinates  used 
by Ma and Bertschinger \cite{Ma_bert}. In this case, $q$ is the comoving momentum of the unperturbed particle and therefore is a constant.  For this choice of momenta, the energy of the particle can be expressed as: $\epsilon = \sqrt{a^2 m_L^2 + q^2}$.

The velocity of the WDM determines its free streaming length scale: $\lambda_{\rm fs} \simeq \int_{a_d}^{a_0} v dt$. The corresponding wavenumber is defined as: $k_{\rm fs}= 2\pi/\lambda_{\rm fs}$. Major contribution to this integral arises from times when the particle is highly relativistic. The comoving free streaming length reaches a maximum at the epoch when  the particle becomes non-relativistic (see e.g. \cite{2006PhR...429..307L} and references therein).  But  as $v\propto 1/a$, free streaming length continues  to be important even after this epoch. In this paper we consider many cases ranging from every early decay   when  the WDM particle is born highly relativistic to late-decay scenarios in which the particle is non-relativistic at the time of WIMP decay. The free streaming scale  determines the impact of the decay product on the growth of  density perturbations. To quantify this effect we define a parameter:
\begin{equation}
    r = m_L/q_L = \frac{2 m_H m_L}{a_d (m_H^2 - m_L^2)}
\label{eq:rdef}
\end{equation}
This parameter denotes the mass to momentum ratio of the WDM. It signifies  the 'coldness' of the WDM particle i.e. in the limit $r\rightarrow \infty$, $m_L \rightarrow m_H$ and the particle is highly non-relativistic at all epochs.  For  $r\rightarrow 0$, $m_H/m_L \rightarrow \infty$. In this case, the particle is highly relativistic at birth and could mimic radiation even at late times.

In the next subsection, We briefly summarize the criteria needed for a viable decaying WIMP model.

\subsection{New parameters}
\label{subsec:param_const}
The decay introduces two more parameters, $a_d$ and $m_L$,  in addition to the self-annihilation cross section  $\sigma$ and  particle mass $m_H$ when the  WIMP is stable. We list below the requirements on these parameters from cosmological observables and give possible implication of this model. 

\begin{enumerate}

 \item We start evolving the abundance of WIMPs before the freeze-out and, assuming s-wave annihilation,   determine the final abundance of WIMPs for a given velocity-weighted cross-section and $m_H$.  The freeze-out occurs at  $x_* \simeq  m_H/T_* \simeq 20$ (e.g. Fig. 4 in \cite{Steigman_2012}), $T_*$, the freeze-out temperature,  corresponds to $a_* \simeq 4.7 \times 10^{-14}$ for a  $100$~GeV WIMP. The final WIMP energy density   is nearly independent of $m_H$ and  can be determined from the relation:  $10^{27} \langle \sigma v \rangle \Omega_{\rm DM} h^2 = 2.1 - 0.3 \log(\Omega_{\rm DM} h^2)$ (Equation~26 of \cite{Steigman_2012}. This issue is discussed in more detail in Appendix~\ref{app:fout}),  where $\sigma$ is the self-annihilation  cross-section of WIMPs  and $v$ is their relative velocity.  The relic abundance $N = \Omega_{\rm DM} \rho_c/m_H$; $\rho_c = 1.879 \times 10^{-29} h^2 \, \rm gm \, cm^{-3}$ is the critical density at the current epoch.  We only consider models for  which the decay occurs after the freeze-out: $a_d > a_*$. In addition, we have the basic kinematic requirement for the decay: $m_H > m_L$.

 \item   For a stable WIMP, $ \langle \sigma  v \rangle  \simeq 3 \times 10^{-26} \, \rm cm^3 sec^{-1}$ to match the observed CDM energy  density. 
The current upper bound on the nucleon-WIMP cross-section from Xenon experiments is $10^{-44} \hbox{--}10^{-47}\, \rm cm^2$ for  a mass range $m_H \simeq 5\hbox{--}1000\, \rm GeV$ \cite{PhysRevLett.121.111302,Ahmed:2010wy,PhysRevLett.118.251302}.  These results  have  ruled out a fraction of parameter-space favoured by supersymmetric extensions of  the standard model (e.g. \cite{2012EPJC...72.2020B,2018RPPh...81f6201R,2019JPhG...46j3003S,2014JHEP...09..081S}). Our proposed model can accommodate a larger range of self-annihilation cross-sections: the current background energy  density of CDM is $m_L N$ if $m_L$ is nonrelativistic at the current epoch. This is smaller by a factor of $m_L/m_H$ as compared to the case of a stable WIMP. In other words, the corresponding increase in relic abundance that results from smaller cross-section  can partly be compensated for by having a lower mass WDM particle produced through decay. This could accommodate both smaller self-annihilation cross sections and a wider    $m_H$--$\sigma$   parameter space. This would also be compatible with current experimental bounds if  $m_L \lesssim 5 \, \rm GeV$, 
as this is the smallest WIMP mass that can be probed by XENON-based experiments \cite{PhysRevLett.121.111302,Ahmed:2010wy,PhysRevLett.118.251302}. 

\item The background WDM density should  match the CDM energy density at the current epoch:
    \begin{equation}
        \label{eq:const_2}\bar{\rho}(a = 1) = N \sqrt{m_L^2 + q_L^2} = Nm_L\sqrt{1 + \left(\frac{1}{r}\right)^2} = \Omega_{\rm DM} \rho_{c}
    \end{equation}
     In the CDM limit,  $r\rightarrow \infty$ and Eq.~(\ref{eq:const_2}) reduces to the usual non-relativistic expression. All the models we study are normalized to the Planck best-fit value of  $\Omega_{\rm DM}$ \cite{aghanim2020}.

\item  The WIMP decay pumps additional radiation into the universe. The fraction of radiation energy contributed by the  massless decay product is:
    \begin{equation}
        \label{eq:const_4} f_{\text{dr}} = \frac{\bar{\rho}_{\text{dr}}}{\bar{\rho}_{\gamma} + \bar{\rho}_{\nu}} = \frac{\Omega_{\rm DM}}{(\Omega_{\gamma}+\Omega_{\nu})\sqrt{1+r^2}}
    \end{equation}
    Here $\bar{\rho}_{\gamma}$  and $\bar{\rho}_{\nu}$ correspond to the background energy density of CMB and three standard model neutrinos, respectively.   In addition, the WDM also contributes to the radiation density before it turns non-relativistic which we discuss in a later section.  Current Planck and big-bang Nucleosynthesis (BBN) results put strong constraints the radiation content of the universe. Planck results determine the relativistic neutrino degrees of  freedom: $2.99 \pm 0.17$ (e.g. see \cite{aghanim2020} for details of joint CMB and BBN ). This limits the additional radiation injection  to be less than 3\% of the contribution of photons and neutrinos.

    \item As we will see in later sections, the most stringent constraints on the decay WIMP model arise from the perturbation analysis of the WDM particle. We defined $r \equiv m_L/q_L$ in the foregoing which determines the 'coldness' of the WDM particles. In addition to the constraint arising from the additional radiation energy density released  in the decay process,  the WDM particles should be sufficiently 'cool' to allow formation of structures at  scales of interest. We consider the perturbation theory  of this particle in the next section.

\end{enumerate}

\section{Perturbation theory of non-thermal WDM}
\label{sec:pert_theory}
As indicated  in the previous section, the main impact of decaying WIMPs can be captured in the parameter $r$, which denotes the coldness of the WDM particle. In this section, we discuss in detail the linear perturbation theory of the decay products of WIMPs---WDM  and radiation. These particles are produced with a phase space distribution given by Eq.~(\ref{eq:psd_WDM}). We work in the Newtonian gauge and  follow the notation of \cite{dodelson} for metric perturbations ($\Psi$ and $\Phi$) and  \cite{Ma_bert} for matter variables.

\subsection{Before the decay: \texorpdfstring{$a < a_d$}{TEXT}}
\begin{figure}[H]
    \centering
    \includegraphics[width = 0.48 \textwidth]{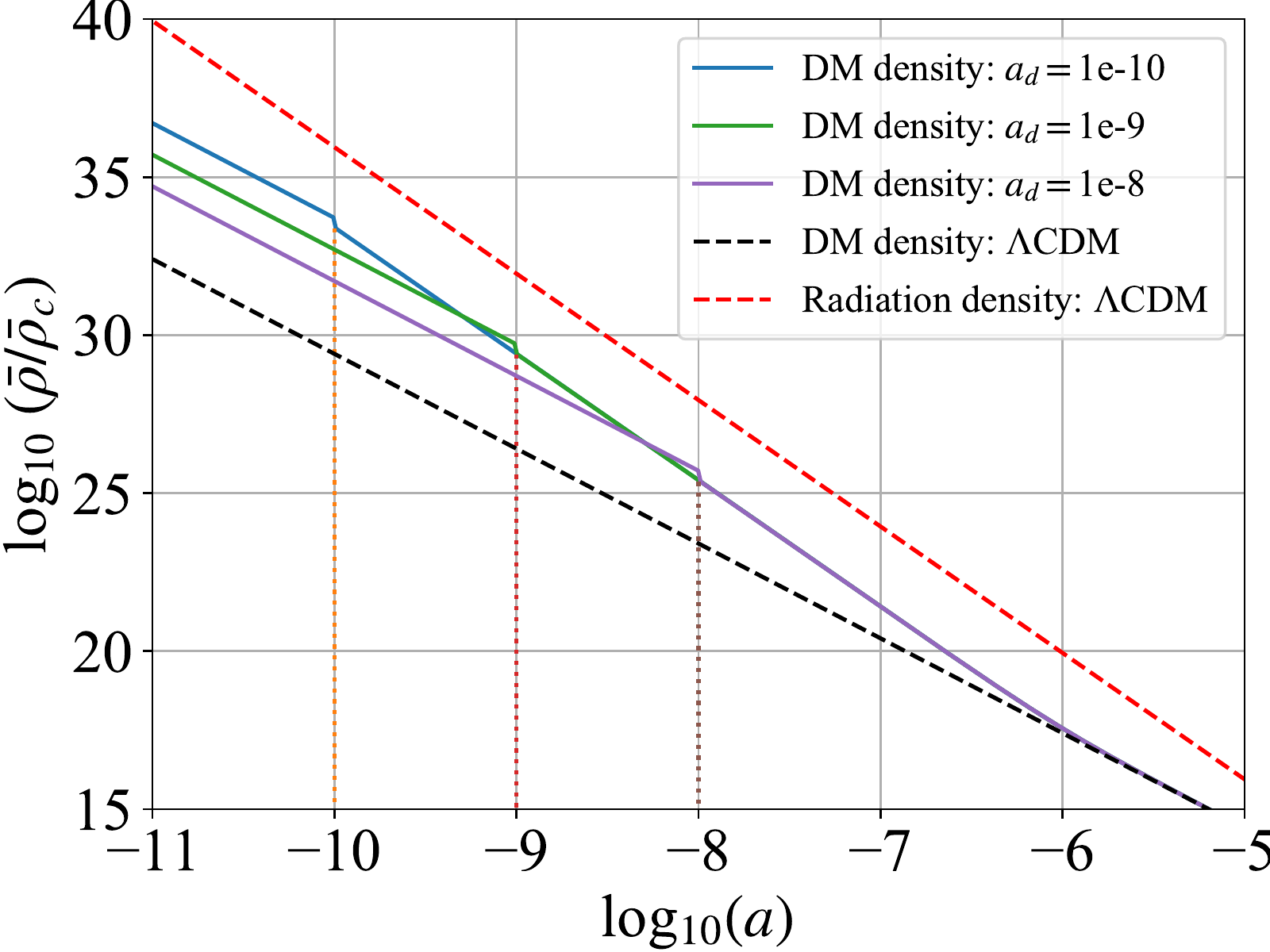}
    \includegraphics[width = 0.48 \textwidth]{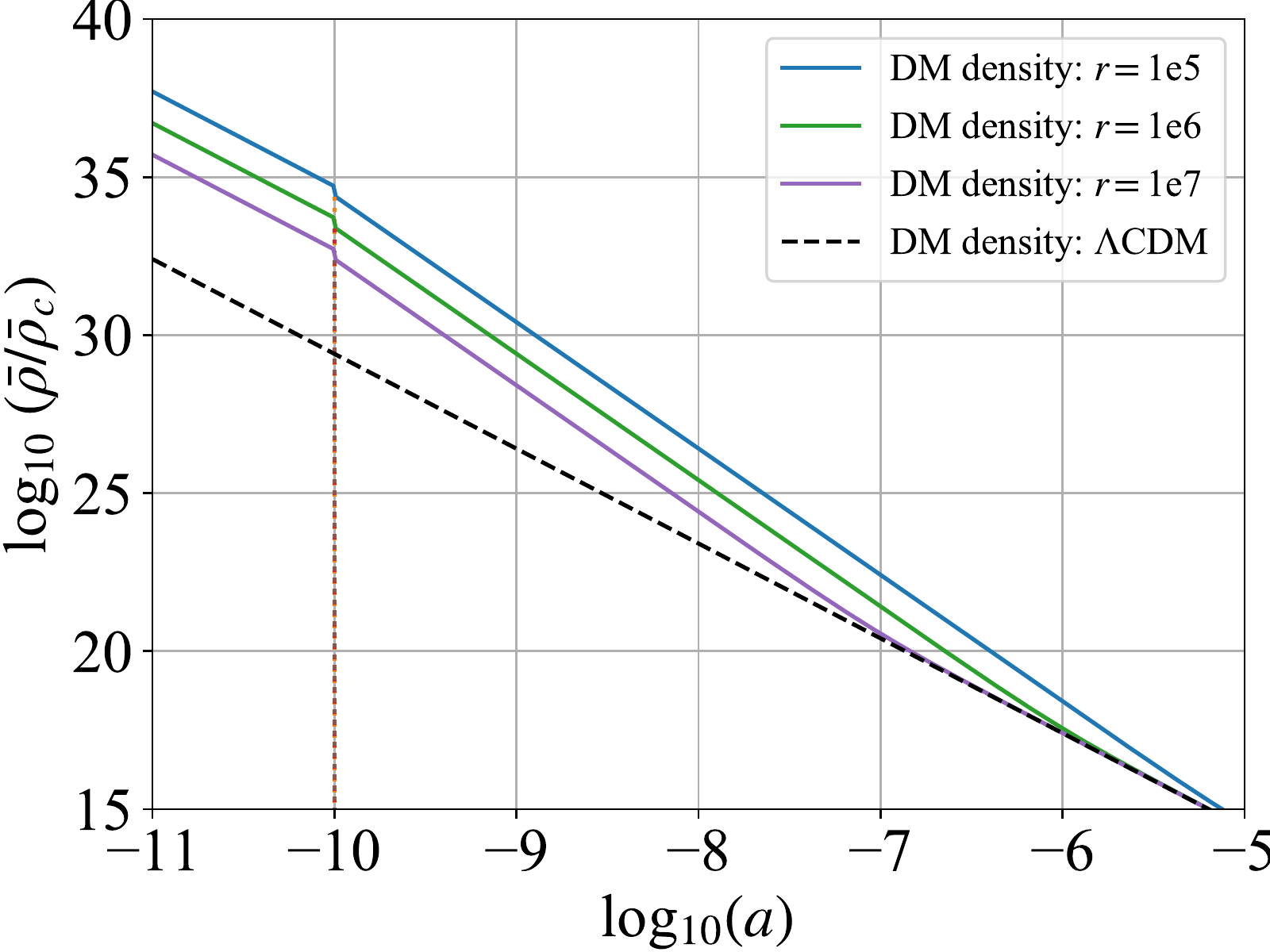}
    \caption{The evolution of the background energy density $\bar{\rho}_{DM}$ is shown for different decaying WIMP models. The radiation and matter energy densities of the $\Lambda$CDM model are shown for comparison.  The left panel shows the  WDM energy densities for three epochs of decay $a_d =  [10^{-10}, 10^{-9}, 10^{-8}]$ and  the same $r = 10^6$. The right panel displays the evolution of enerfy densities for a fixed $a_d$ and different $r = [10^5, 10^6, 10^7]$. The  energy density of WIMPs, $\bar{\rho}$,   falls as $1/a^3$  prior to decay (Eq.~(\ref{eq:0_ord_quant_CDM})) After the decay, it follows WDM density (\eqref{eq:0_ord_quant_WDM}) which is independent of $a_d$ and therefore all the  curves (in both panels) converge at low redshifts. In the limit $r \gg 1$ and large $a$, Eq.~\eqref{eq:0_ord_quant_WDM} reduces to the CDM component, resulting in all the curves merging with the standard CDM case at low redshifts. The radiation component of the WIMP decay product is not shown. There is a step-like decrement at $a_d$, which corresponds to the decay of all the WIMPs into lighter WDM and radiation component. In a realistic scenario, the decay should be over a time interval leading to a continuous change in density. For our purposes, we assume that the width of this interval is negligible as compared to the cosmological timescale.}
    \label{fig:rho_dm}
\end{figure}

After the freeze-out,  WIMPs behave as  cold dark matter with  comoving relic abundance $N$ as determined by their self-annihilation  cross-section $\sigma$. The background energy density and pressure are  given by:
\begin{eqnarray}
    \label{eq:0_ord_quant_CDM} \bar{\rho}& = &\frac{m_H N}{a^3} =  \frac{\Omega_{\rm DM} \rho_{c}}{a^3} \frac{m_H}{\sqrt{m_L^2 +q_L^2}}=  \frac{\Omega_{\rm DM} \rho_{c}}{a^3} \frac{a_d r^2}{(\sqrt{1+(a_d r)^2} - 1) \sqrt{1 + r^2}} \nonumber  \\
     \bar{P} & =&  0 
\end{eqnarray}
We note that the initial WIMP background density  in our case is larger than the  $\Lambda$CDM model by a factor  $m_H/\sqrt{m_L^2 + q_L^2}$. This is  because the decay products of WIMPs could  behave as radiation for long periods after the decay (Figure~\ref{fig:rho_dm}) and, therefore, to satisfy the density constraint (Eq.~\eqref{eq:const_2}), the initial WIMP density has to be higher than for the case of a stable WIMP  assumed in the  $\Lambda$CDM model. 

In this case, the Boltzmann equations for density contrast $\delta$ and divergence component of the bulk velocity $\theta = ik^i v_i$ reduce to the usual CDM case (Equations~(43) in \cite{Ma_bert}; for details of obtaining fluid equations from Boltzmann equations see e.g. \cite{dodelson}):
\begin{equation}
    \label{eq:Boltz_CDM} \dot{\delta} = - \theta - 3 \dot{\Phi}
    \hspace{0.5 cm}\dot{\theta} = - \frac{\dot{a}}{a} \theta + k^2\Psi
\end{equation}
And the initial conditions for these variables are (Equations~(97), (98) in \cite{Ma_bert}):
\begin{equation}
    \label{eq:init_CDM}\delta = -\frac{3}{2}\Psi
    \hspace{0.5 cm} \theta = \frac{k^2\eta}{2}\Psi
\end{equation}

\subsection{After the decay: \texorpdfstring{$a > a_d$}{TEXT}}
The WIMP decays into WDM and radiation. The WDM produced at decay is not necessarily either highly relativistic or non-relativistic. In most cases of interest to us in the paper, the WDM is born highly relativistic and make a transition to a non-relativistic particle   during its evolution. In the general case,  we cannot separate the $a$ and $q$ dependence in  the WDM  energy $\epsilon = \sqrt{q^2 + a^2m_L^2}$,  and one has to  numerically integrate over comoving momenta $q$ in the equations of Boltzmann hierarchy.  However, unlike the usual case of HDM/WDM particles for which the 
phase space distribution of particles could be Fermi-Dirac distribution,  the phase space distribution function in our cases is a delta function which allows us to analytically integrate over the momenta to obtain both  the background density $\bar{\rho}$ and pressure $\bar{P}$ as well as the first order quantities: overdensity $\delta$, bulk velocity $\theta$, and shear stress $\sigma$.  

 To obtain background quantities, we plug in the WDM distribution function Eq.~\eqref{eq:psd_WDM} into Equation~(52) of \cite{Ma_bert}, replace the relic abundance $N$ in favour of $\Omega_{\rm DM}$ using \eqref{eq:const_2} and $r = m_L/q_L$. This yields:
\begin{equation}
    \label{eq:0_ord_quant_WDM} \bar{\rho} = \frac{\Omega_{\rm DM}\rho_{c}}{a^4} \sqrt{\frac{1+r^2 a^2}{1 + r^2}} \hspace{0.5 cm} \bar{P} = \frac{\Omega_{DM}\rho_{c}}{3a^4} \frac{1}{\sqrt{(1+r^2a^2)(1+r^2)}}
\end{equation}

Similarly, using Equation~(55) in \cite{Ma_bert}, one can calculate the 1st order quantities:
\begin{eqnarray}
    \bar{\rho} \delta &= & \Delta_0\frac{\Omega_{\rm DM}\rho_{c}}{a^4} \sqrt{\frac{1+r^2 a^2}{1 + r^2}} \nonumber \\
    (\bar{\rho} + \bar{P}) \theta &= &k\Delta_1\frac{\Omega_{\rm DM}\rho_{c}}{a^4} \frac{1}{\sqrt{1 + r^2}}  \nonumber\\
    (\bar{\rho} + \bar{P}) \sigma &= & \Delta_2\frac{\Omega_{\rm DM}\rho_{c}}{a^4} \frac{2}{3\sqrt{(1+r^2a^2)(1+r^2)}}
    \label{eq:1_ord_quant_WDM}
\end{eqnarray}

For the Boltzmann equations (Equation~(57) in \cite{Ma_bert}),   $q$-dependence in the  equations is integrated using the WDM distribution function over with $\int \epsilon f_0 q^2 dq, \int q f_0 q^2 dq,$ $\int q f_0 q^2 dq$, respectively. Using Eq.~\eqref{eq:1_ord_quant_WDM}, we  get:
\begin{align}
    \label{eq:Boltz_WDM} \dot{\delta} &= - \frac{4 + 3a^2r^2}{1 + a^2r^2} \Big[ \frac{1}{3}\theta + \dot{\Phi} \Big] \nonumber \\
    \dot{\theta} &= - \frac{\dot{a}}{a} \frac{(a^2r^2)(2+3a^2r^2)}{(1+a^2r^2)(4+3a^2r^2)}\theta + k^2 \Big[ \frac{\delta}{4+3a^2r^2} - \sigma \Big] + k^2\Psi \nonumber \\
    \dot{\Delta_l} &= \frac{k}{2l+1} \frac{1}{\sqrt{1 + a^2r^2}}\big( l\Delta_{l-1} - (l+1)\Delta_{l+1} \big) \hspace{0.5 cm} \text{for  }l \geq 2
\end{align}
Here $\Delta_l$ can be defined in terms of the perturbed phase space distribution
function $\chi(k,\hat n,q,\eta)$ (for details see \cite{Ma_bert}):
\begin{equation}
\Delta_l(k,\eta) = \frac{\int dq \chi_l(k,q,\eta) f_0 q^3}{\int dq q^3 f_0}
\end{equation}
with 
\begin{equation}
\chi_l(k,q,\eta)  = \int d\theta \chi(k,\hat n,q,\eta) P_l(\theta)
\end{equation}
Here $\theta = \hat k.\hat n$ is the angle between the unit vectors of the  Fourier mode and the  particle momentum.

Finally, we use the following prescription to truncate  the Boltzmann heirarchy (Equation~(58) in \cite{Ma_bert}):
\begin{equation}
    \label{eq:terminal_Boltz_WDM} \Delta_{l_{max}+1} = \frac{2l_{max}+1}{k\eta}\sqrt{1 + a^2r^2} \Delta_{l_{max}} -  \Delta_{l_{max}-1}
\end{equation}
We take the initial conditions for WDM to match those of the WIMP at the time of decay. It should  be noted that the entire set of perturbative equations depend only on two parameters: the mass-momentum ratio at the  decay $r$ and the current DM density $\Omega_{\rm \rm DM}$. 

 The dynamics of WDM perturbations  also depend weakly on the epoch of decay $a_d$ through the background WIMP density Eq.~(\eqref{eq:0_ord_quant_CDM}) and the switch condition from CDM to WDM equations at $a = a_d$. However, for the parameter space of interest to us, this dependence is negligible.
 
Next, we consider the equations corresponding  to the radiation component of  the decay. This component has the same  phase space distribution function as the WDM (Eq.~\eqref{eq:psd_WDM}). For radiation, the energy $\epsilon = q$.  We use Equation~(44) of \cite{Ma_bert} to obtain the background quantities:
\begin{equation}
    \label{eq:0_ord_quant_rad} \bar{\rho}_{\rm dr} = 3\bar{P}_{\rm dr} = \frac{\Omega_{\text{dr}}\rho_{c}}{a^4} \hspace{0.5 cm} \Omega_{\text{dr}} = \frac{\Omega_{\rm DM}}{\sqrt{1 + r^2}}
\end{equation}
Similarly, Equation~(47) of \cite{Ma_bert} yields the  1st order quantities:
\begin{equation}
    \label{eq:1_ord_quant_rad} \delta = \Delta_0 \hspace{0.5 cm}
    \theta = \frac{3}{4} k \Delta_1 \hspace{0.5 cm}
    \sigma = \frac{1}{2}\Delta_2
\end{equation}
And the $q$-dependence in the Boltzmann equations (Equations~(49) of \cite{Ma_bert}) can be integrated over since $\epsilon$ does not depend on $a$. It should  noted that even though the distribution function of this additional radiation component is different from that of massless neutrinos, the resulting equations are the same. More generally, so long as the $q$ and $a$ dependence of the energy $\epsilon$ can be separated i.e. in highly relativistic $\epsilon = q$ or completely non-relativistic $\epsilon = am$ case, the $q$-dependence can be removed by integrating analytically. The relevant equations are: 
\begin{eqnarray}
     \dot{\delta} &= & -\frac{4}{3}\theta - 4 \dot{\Phi} \nonumber \\ 
    \dot{\theta} &= & k^2 \Big[ \frac{\delta}{4} - \sigma \Big] + k^2\Psi \nonumber \\
    \dot{\Delta_l} & =  & \frac{k}{2l+1} \big( l\Delta_{l-1} - (l+1)\Delta_{l+1} \big)  \, \, \, \hbox{for } l \geq 2
    \label{eq:Boltz_rad}
\end{eqnarray}
We use the following condition  to truncate the Boltzmann heirarchy  (Equation~(51) in \cite{Ma_bert}):
\begin{equation}
    \label{eq:terminal_Boltz_rad} \Delta_{l_{max}+1} = \frac{2l_{max}+1}{k\eta} \Delta_{l_{max}} -  \Delta_{l_{max}-1}
\end{equation}
As for the WDM particle, the initial conditions for $\delta, \theta$, and $\sigma$ of this decay radiation are the same as that of WIMP at decay. This radiation component evolves as an independent species. Its contribution to the total radiation component of the universe is determined by $\Omega_{\rm DM}$ and $r$. 

As anticipated in section~\ref{subsec:param_const}, the dynamics of perturbations in our case is solely determined 
by the value of $r$. The  joining conditions at $a = a_d$ cause a very weak dependence of the results on the value of $a_d$.
One can verify that in the limit $r\rightarrow \infty$ (non-relativistic limit),  Eqs.~(\ref{eq:Boltz_WDM})  reduce to  CDM equations (Eqs.~\eqref{eq:Boltz_CDM}).  In this case,  the equations reduce to non-interacting fluid equations (continuity and Euler equations). All the higher order moments (Eqs.~(\ref{eq:terminal_Boltz_WDM})) vanish as they are suppressed by powers of  $q/\epsilon$.  
On the other hand,  for $r = 0$ (relativistic limit),  Eqs.~(\ref{eq:Boltz_WDM}) reduce to  Eqs.~(\ref{eq:Boltz_rad}), which give the evolution of perturbations of massless particles.
 It is also readily checked that the  background quantities and matter variables (Eqs.~(\ref{eq:0_ord_quant_WDM})  and~(\ref{eq:1_ord_quant_WDM})) also reduce to expressions appropriate for CDM and massless particles in these limits.

\subsubsection{Matter-radiation equality}
\label{sec:submreq}
As the matter-radiation equality is determined with high precision by the Planck CMB data ($z_{\rm eq} = 3387 \pm 21$ \cite{aghanim2020}), we briefly discuss how this epoch is altered in our case. Both the decay products of WIMPs contribute to radiation energy density in the universe.  
The  WDM contributes to both   the radiation  and matter energy density:  $\bar{\rho}_{r} = 3\bar{P}$  (Eq.~(\ref{eq:0_ord_quant_WDM})) and  $\bar{\rho}_m = \bar{\rho} - 3\bar{P}$. Therefore, the total radiation density receives contribution from the decay radiation as well as the WDM in addition to photons and massless neutrinos, because of which the matter-radiation equality shifts closer to the current epoch as compared to the CDM case. At matter-radiation equality, $\bar{\rho}_{m} + \bar{\rho}_b = \bar{\rho}_{r} + \bar{\rho}_{\text{dr}} + \bar{\rho}_{\gamma} + \bar{\rho}_{\nu}$;  $\bar{\rho}_b$ is the background baryonic density. Using Eqs.~\eqref{eq:0_ord_quant_WDM} and~\eqref{eq:0_ord_quant_rad}, the scale factor $a_{\rm eq}$ can be determined from the condition:
\begin{equation}
    \label{eq:a_eq}\Omega_{\rm DM} \Big[ \sqrt{\frac{1 + a_{eq}^2r^2}{1+r^2}} - \frac{2}{\sqrt{(1+a_{eq}^2r^2)(1+r^2}} - \frac{1}{\sqrt{1+r^2}}\Big] + \Omega_b a_{eq} - (\Omega_{\gamma} + \Omega_{\nu}) = 0
\end{equation}

\subsection{CLASS implementation}
Cosmic Linear Anisotropy Solving System \cite{CLASS_2011} is one of the standard packages used for numerically solving coupled Einstein and Boltzmann equations for multiple coupled fluids in the context of  cosmological perturbation theory. We modify the CLASS codes to add the relevant equations for the decay products---WDM and radiation---of the WIMP. The following major changes  were made to the existing $\Lambda$CDM model already implemented in CLASS:
\begin{enumerate}
    \item We added Eqs.~(\ref{eq:0_ord_quant_CDM})--(\ref{eq:terminal_Boltz_rad}) into relevant places in CLASS codes. The default parameter corresponding to the cold dark energy density  $\Omega_{\rm CDM}$ is set to zero. Instead three new parameters $\Omega_{\rm DM}$, $r$,   and $a_d$ are used to quantify the WDM and the decay  radiation ($\Omega_{\rm dr}$ is computed from  (Eq.~(\ref{eq:0_ord_quant_rad})). As noted above, $\Omega_{\rm DM}$ is normalized to $\Omega_{\rm CDM}$ at
    $a=1$. As shown  in the foregoing, these parameters can be expressed in  terms of parameters: $m_H$, $m_L$, $\langle \sigma v \rangle$, and $a_d$. 
    
    \item The background density and pressure for the dark matter  are calculated using CDM equations Eq.~\eqref{eq:0_ord_quant_CDM} for $a < a_d$ and WDM and radiation relations for $a >  a_d$ (Eqs.~\eqref{eq:0_ord_quant_WDM} and~\eqref{eq:0_ord_quant_rad}).

    \item For the perturbed components,  Eqs.~\eqref{eq:Boltz_CDM} are solved with initial conditions given by Eq.~(\ref{eq:init_CDM})  for $a<a_d$. At $a = a_d$, we switch to Boltzmann equations for WDM (Eq.~\eqref{eq:Boltz_WDM}) and radiation (Eq.~\eqref{eq:Boltz_rad}). 
    The initial conditions at $a=a_d$ follow from the value of $\delta$ and $\theta$ of WIMPs at that time. Notice that as 
    $\sigma = 0$ for WIMPs, both WDM and radiation inherit this initial condition. This situation is akin to the compensated mode
    when both massless neutrinos and primordial magnetic fields are present (e.g. \cite{2010PhRvD..81d3517S}). 

    \item An alternative way to incorporate WDM component in CLASS is through the non-cold dark matter option (using keyword {\it ncdm}).
    This allows one to input a distribution function different from the default Fermi-Dirac distribution used for massive neutrinos. 
    We input the following distribution function to approximate the delta function:
    \begin{equation*}
    f_0(q) = \frac{N}{4\pi q^2} \lim_{\alpha \to 0} \frac{1}{\sqrt{2\pi \alpha}} \exp{\Big[-\frac{(q - a_d m_H/2)^2}{2\alpha}\Big]}
\end{equation*}
We varied $\alpha$ and studied its impact on the output. The procedure converges and yields the same results  as obtained by direct inclusion of relevant equations in the code.  We prefer the  inclusion of relevant equations in CLASS over the {\it ncdm} option for the following reasons: 
\begin{enumerate}
    \item CLASS uses a number of approximation schemes---tight-coupling, ncdm fluid, free streaming, and ultra relativistic, to speed up the computation. For each of these approximations, CLASS computes the intervals within which they are valid. For the range of parameter space considered in our study, we find it difficult to match these conditions across the boundaries of their validity. Improper matching  could result in  sharp breaks and erroneous values at large scales in the matter power spectrum. 
    \item This approach still requires integration of the Boltzmann hierarchy over a grid of $q$ values. Thus, the computation cost is significantly increased as nearly 20 times (for 20 $q$-bins) the number of equations have to be  solved for each time step. The run time for such cases is about $20$ seconds.  However,  by solving the Boltzmann equations after integrating over $q$ brings the run time down to $0.5$ seconds.
    \item The decay product corresponding to massless particles cannot be incorporated using  only the {\it ncdm} option. As the fraction contributed by the additional radiation component is less than 0.3\% of the total radiation for models of interest, its impact is  negligible.  It  also allows us to  compare the two methods.
\end{enumerate}
More details of the implementation of our model in CLASS are given in the Appendix.

\end{enumerate}

\section{Results}
\label{sec:resall}
In addition to CLASS implementation, we developed a Python code to numerically compute the matter power spectrum for a mutliple component fluid consisting of, in addition to WIMP and its decay products, photons, baryons, and massless neutrinos.  For 
both our codes and CLASS runs we chose the following models/settings: (a) spatially flat universe with dark energy assumed to be  cosmological constant, (b) adiabatic initial condition, (c) evolution from scale factor $a = 10^{-14}$ to $a = 1$.
The other settings were chosen as  the default settings for CLASS. The matter power spectra we obtain from our  codes are  in excellent agreement with CLASS results. 

In Figure~\ref{fig:rho_dm}, we show the background evolution of the energy density of WDM for a range of decay times, $a_d$ and $r$. All the models shown are normalized such that the energy
density matches the observed CDM energy density at the current epoch.

The main difference between  decaying WIMP and the $\Lambda$CDM model emanates from two distinct reasons: (a) matter-radiation equality: From Eqs.~\eqref{eq:0_ord_quant_WDM} and~\eqref{eq:0_ord_quant_rad}, it is clear that both the decay products of WIMPs (WDM and radiation) contribute to radiation energy density. This can delay the matter radiation equality. For $r = \{10^5, 10^6\}$, we obtain
$a_{\rm eq} = \{3.024\times 10^{-4},  2.943\times 10^{-4}\}$, respectively. This leaves a detectable signature on both CMB and galaxy data. The joint analysis of Planck CMB and galaxy  BAO data yields: $z_{\rm eq} = 1/a_{\rm eq} = 3387 \pm 21$ \cite{aghanim2020}. (b) free-streaming of WDM: The free streaming of WDM hinders the formation of structures at sub-horizon scales. While this phenomenon effects a range of scales at different times, its  impact 
can be approximately captured by  the free-streaming length scale $k_{\rm fs}$ defined in the foregoing. We note that $k_{\rm fs} =\{0.094, 0.92\} \, \rm h Mpc^{-1}$ for $r = \{10^5, 10^6\}$.

\begin{figure}[H]
    \centering
    \includegraphics[width = 0.49\textwidth]{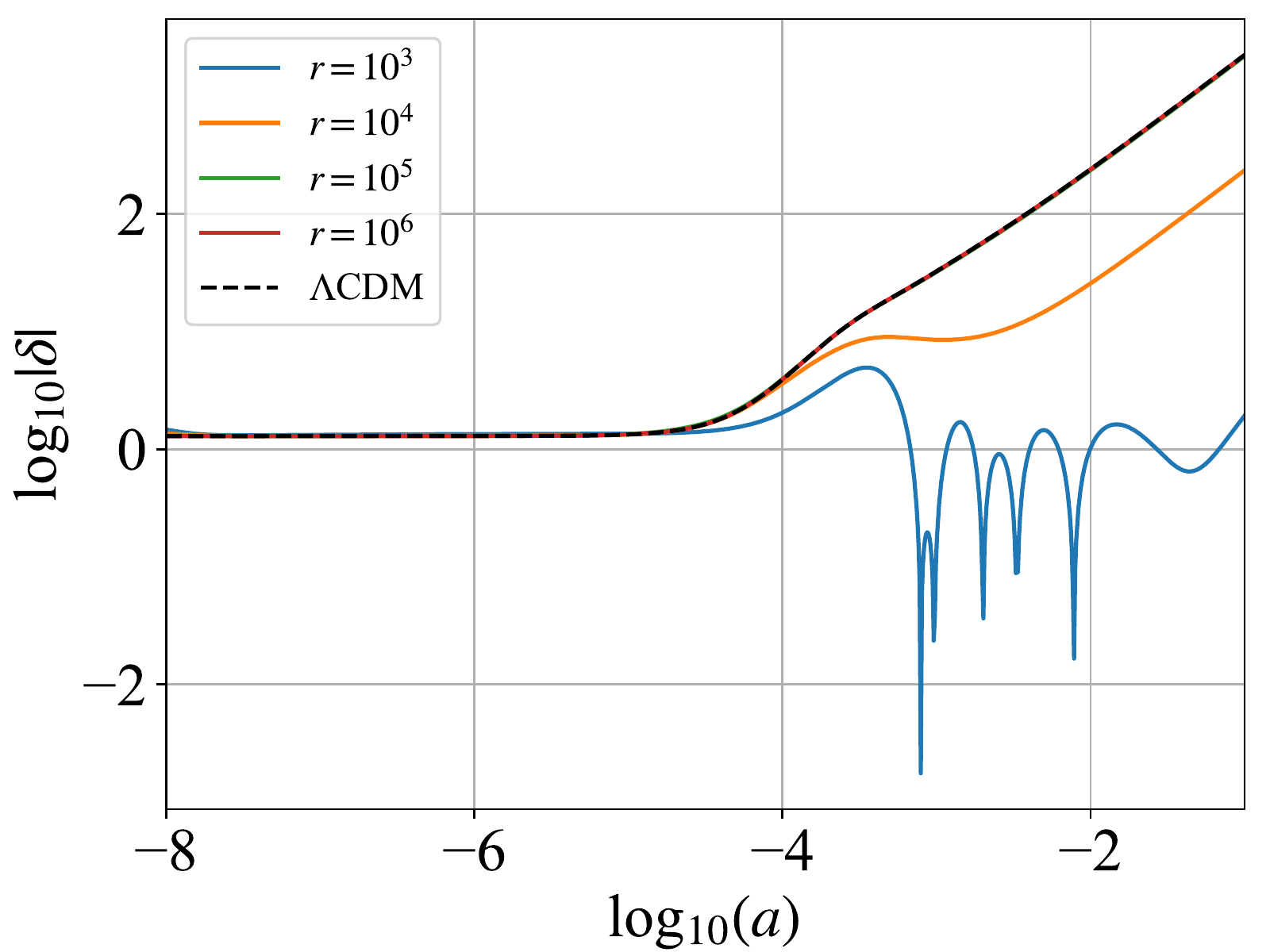}
    \includegraphics[width = 0.49\textwidth]{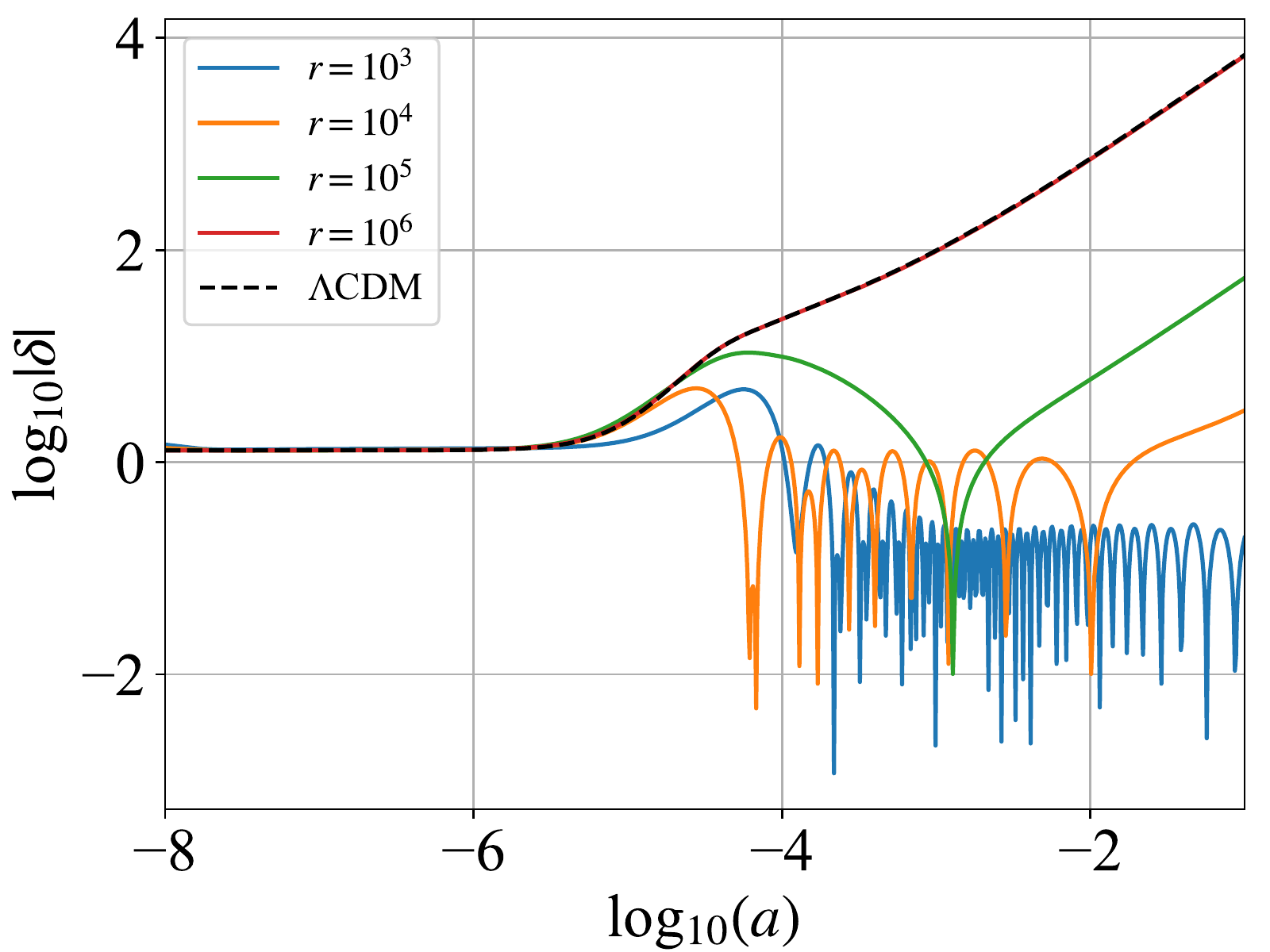}
    \caption{The evolution of overdensity is shown for two scales $k = 0.1$ h Mpc$^{-1}$ (left panel) and $k = 0.6$ h Mpc$^{-1}$ (right panel). Evolution of the same scales is also shown for the $\Lambda$CDM model for comparison. The epoch of decay  $a_d =  10^{-10}$.}
    \label{fig:results_WDM}
\end{figure}

In Figure~\ref{fig:results_WDM}, we show the density evolution for two scales for different values of $r$. The $\Lambda$CDM model
is also shown for comparison. The figure allows us to verify our understanding of the expected behaviour as $r$ changes. As noted above,
the WDM is 'warmer' for small $r$. For $r=10^3$, the perturbations oscillate and decay after horizon entry. This behaviour is akin 
to a massless neutrino. As $r$ is increased the particle becomes 'cooler' and  the power spectra approach  the $\Lambda$CDM model.
The evolution of the mode is 
indistinguishable from the standard model for  $k = 0.1 \, \rm h Mpc^{-1}$  and $r> 10^5$.
As both decaying WIMP and $\Lambda$CDM models have identical evolution at super-horizon scales, the convergence towards the standard case  is also more prominent for scales that enter the  horizon later.   For instance, the $k = 0.6 \, \rm h Mpc^{-1}$ converges for larger value of $r$ as compared to the $k = 0.1 \, \rm h Mpc^{-1}$ mode because this mode enters the horizon earlier. 
As the particle is hotter at earlier epochs, the density perturbations on smaller scales deviate  more significantly from the $\Lambda$CDM model.

\begin{figure}[H]
    \centering
    \includegraphics[width = 0.49\textwidth]{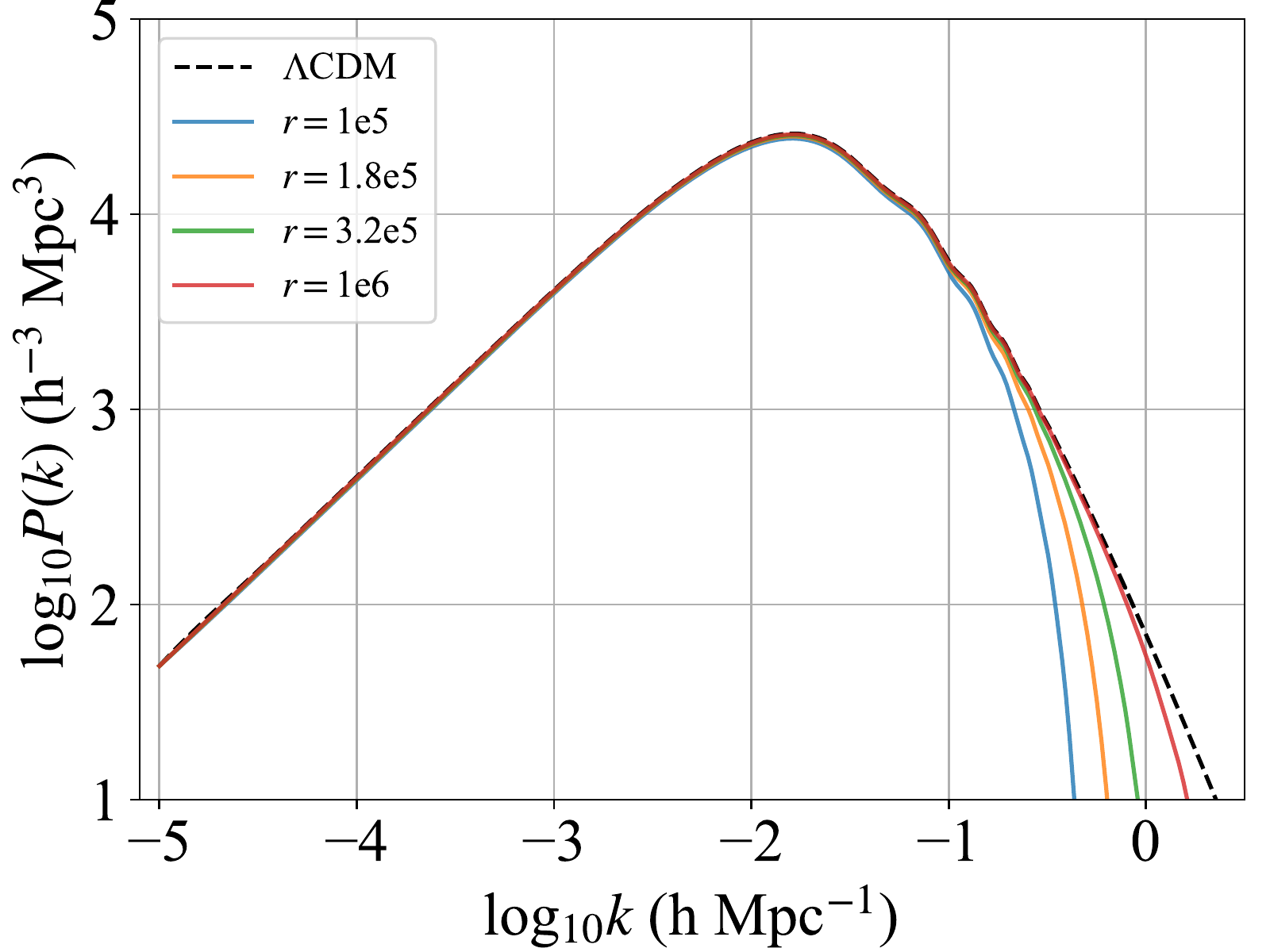}
    \includegraphics[width = 0.49\textwidth]{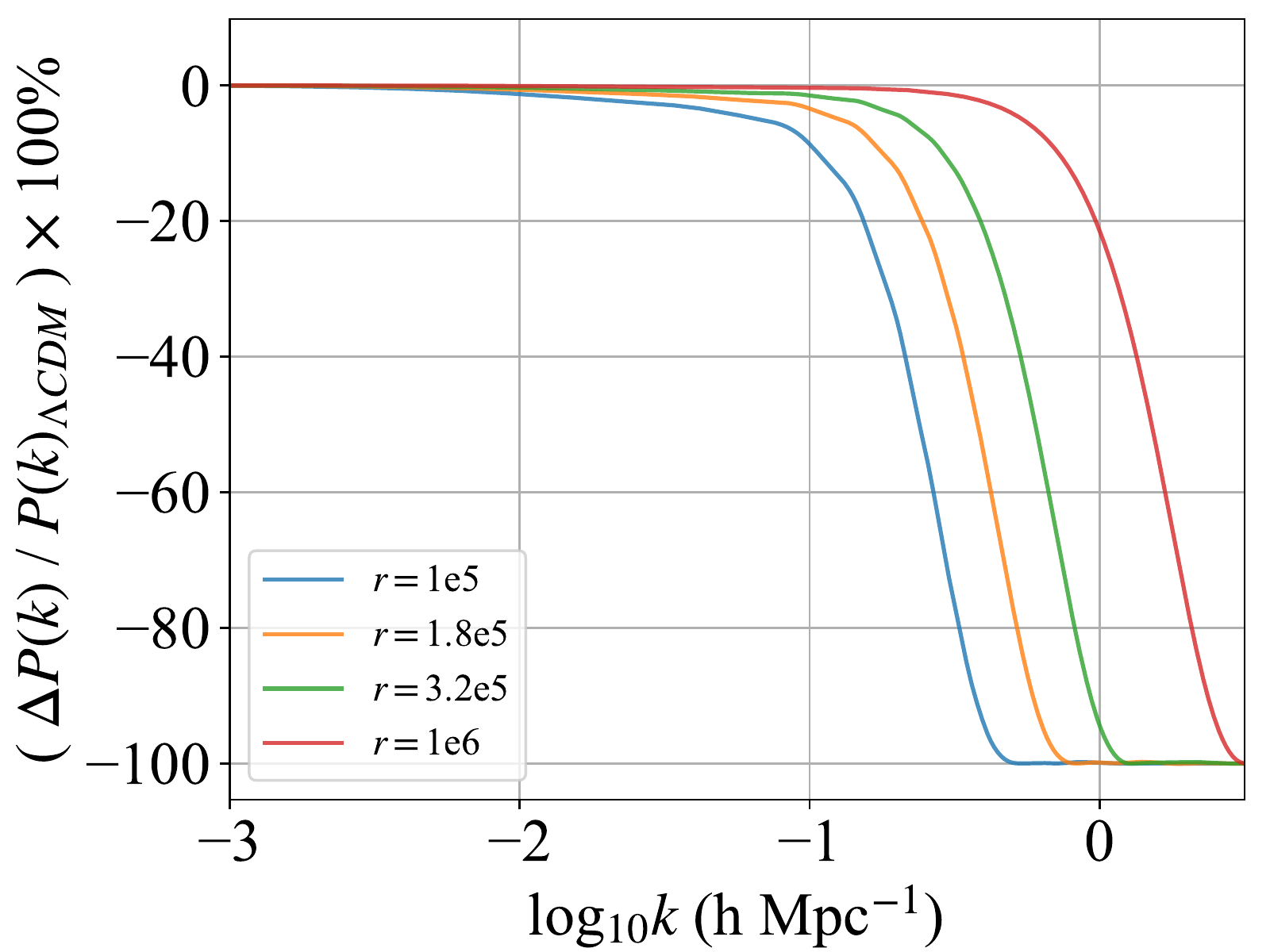}
    \caption{ Left Panel: The linear matter power spectra at $z = 0$ for different values of $r$ are shown along with the  $\Lambda$CDM model. Right Panel: The percentage difference between the $\Lambda$CDM   and  decaying WIMP matter power spectra is shown. }
    \label{fig:results_pscmb}
\end{figure}

\begin{figure}[H]
    \centering
    \includegraphics[width = 0.49\textwidth]{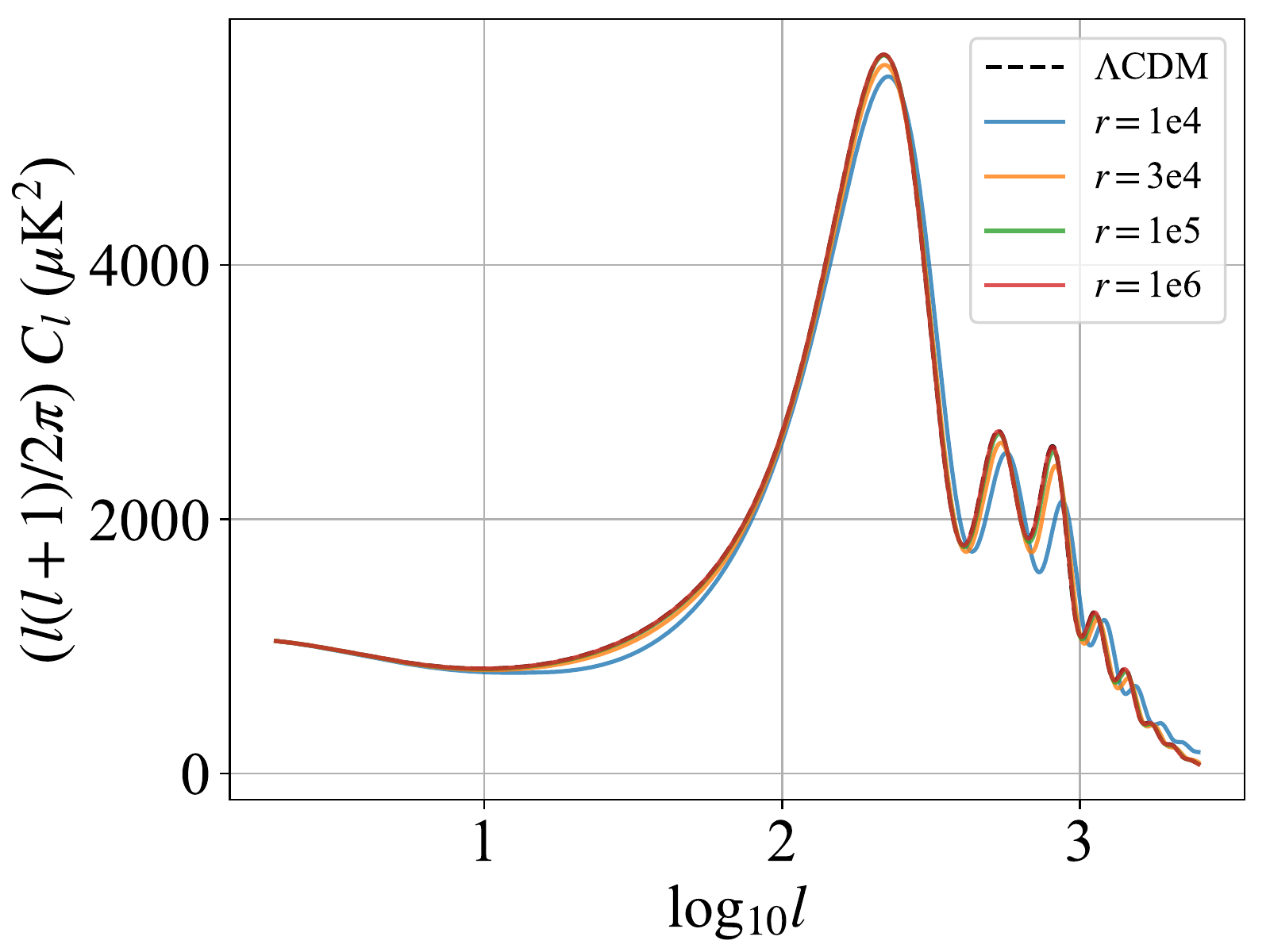}
    \includegraphics[width = 0.49\textwidth]{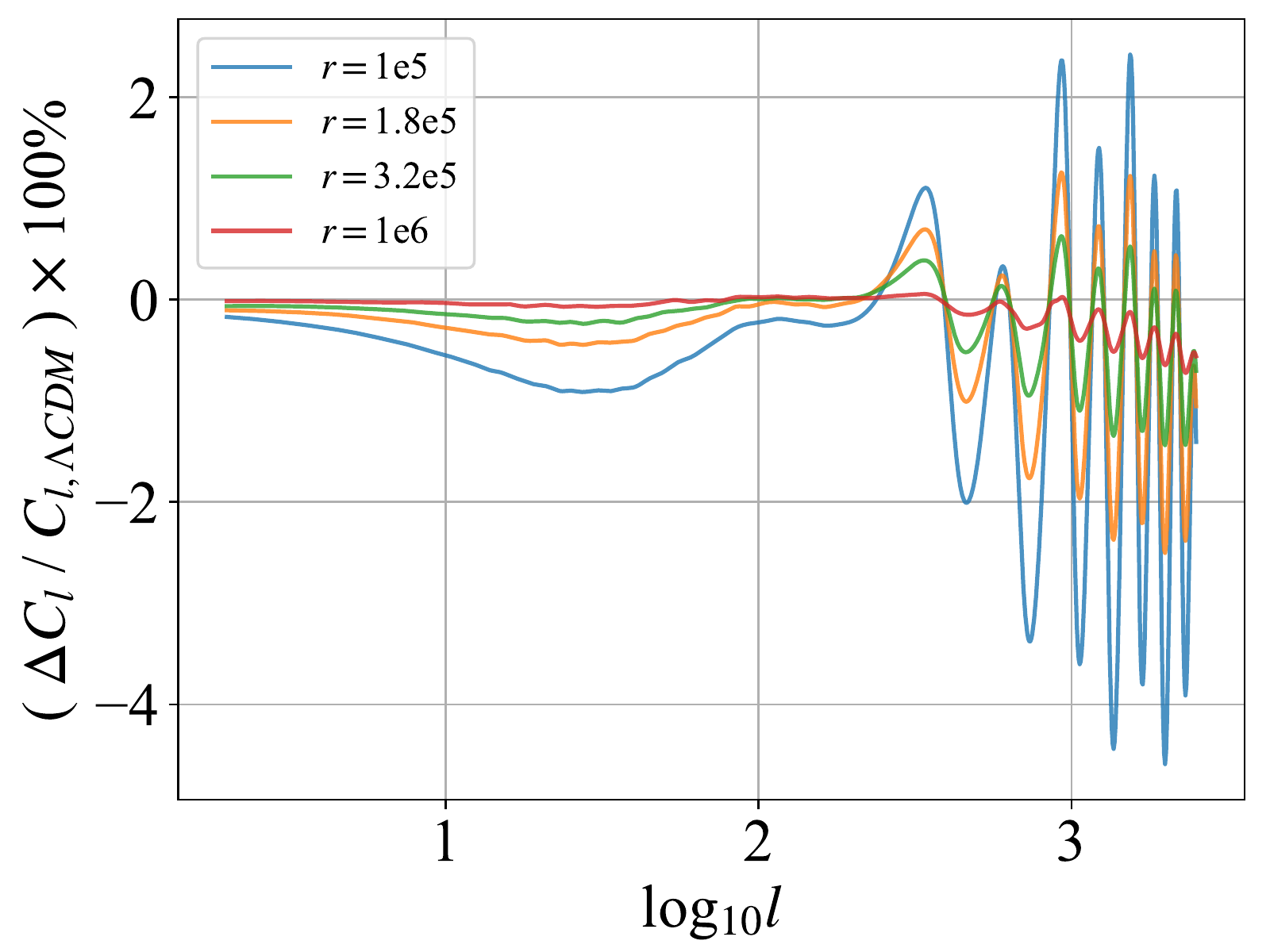}
    \caption{ Left Panel: The temperature-temperature (TT) angular power spectrum  is for CMB for different values of $r$ are shown along with the  $\Lambda$CDM model. Right Panel: The percentage difference between the $\Lambda$CDM  and  decaying WIMP models is shown for CMB TT angular power spectrum. }
    \label{fig:results_pscmb1}
\end{figure}

In Figure~\ref{fig:results_pscmb} we compare  the linear  matter power spectra for different $r$. The results are in line with 
the evolution of density perturbations shown in Figure~\ref{fig:results_WDM}: the power on small scales is suppressed as these scale
enter the horizon at earlier times when the particle is hotter. In the right panel we show the percentage difference between
decaying WDM and $\Lambda$CDM models, which show the scales at which the power is suppressed more clearly. The approximate scales of suppression can  be gauged
from the free-streaming scales: For $r = 10^5$,  $k_{\rm fs} \simeq 0.09  \, \rm h Mpc^{-1}$ while it is nearly an order of magnitude larger for $r = 10^6$. These scales give the approximate point of departure between the $\Lambda$CDM and WDM  linear matter power spectra, for corresponding values of $r$, in Figure~\ref{fig:results_WDM}.

The impact of the altered dynamics  on the CMB angular 
power spectrum is shown in Figure~\ref{fig:results_pscmb1}. The difference between our models and the $\Lambda$CDM  model arises owing to:
(a) the suppression of matter power: this results in a reduction in the amplitude of CMB peaks, (b) increase in the expansion rate $H$: additional radiation
energy  increases the expansion rate, which causes a decrease in the sound horizon $r_s \simeq 1/3\int d\eta/(1+R)$  ($R = 3\rho_b/(4\rho_\gamma)$) at the last scattering surface. As the angular scales of  CMB peaks
correspond to the  multiples of the sound horizon, this causes the peaks to shift to smaller angular scales or larger $\ell$.  (a) and (b) together are responsible for the observed difference at $\ell \gtrsim 200$ shown in the right panel  of Figure~\ref{fig:results_pscmb1}. 
In addition,  an enhancement in the integrated Sachs-Wolfe effect owing to additional radiation component at the last scattering surface  causes the angular power spectra to differ at smaller $\ell$. 
As expected, the models approach the $\Lambda$CDM model
as $r$ increases. 

 Before undertaking detailed   comparison of our models with  data in the next section, we briefly discuss how cosmological observables might constrain parameters in our model.  In our analysis, there are four parameters---$m_H$, $m_L$, $a_d$, and $\langle \sigma v \rangle$----and two constraints
on $\Omega_{\rm DM}$ and  $r$ from cosmological observables (we do not list $\Omega_{\rm dr}$ as a separate parameter, as it 
can be derived from these two parameters (Eq.~(\ref{eq:0_ord_quant_rad}))).  For the models we consider 
the freeze out abundance of WIMPs, $N \propto 1/ (m_H\langle \sigma v \rangle)$ could be much larger  than the abundance in the stable WIMP model (Figure~\ref{fig:rho_dm}), which requires $m_L$ to be smaller than $m_H$ to match the current mass density. The cosmological  dark matter mass density $\Omega_{\rm DM}h^2 \propto m_L/ (m_H\langle \sigma v \rangle)$ and 
fixing the mass density to its best-fit value allows us to eliminate one parameter.  The CMB anisotropies and galaxy clustering data puts a lower bound on 
$r \simeq 2m_L/(a_d m_H)$ (Eq.~(\ref{eq:rdef})). This constraint enables the elimination of another parameter which reduces the 
allowed parameter space from four to two. It follows from the two constraints that the  allowed region is a region  bounded by two surfaces---(i) $\langle \sigma v \rangle / a_d > C_1$ (ii) $m_L / (m_H a_d) > C_2$, where $C_1, C_2$ are constants.


\subsection{Data analysis}
\label{sec:data_analysis}
We compare our models based on linear perturbation theory against the available  CMB and galaxy clustering data. In particular, we use Planck 2018  likelihood data of  angular power spectra  of temperature, polarisation (E mode) and lensing potential \cite{aghanim2020}. From SDSS data, we use the likelihood data of two-point correlation function of the  distribution of Luminous Red Galaxies \cite{BOSS_bao}. The CMB data estimates the angular power spectra for angular modes $\ell \lesssim 2500$, which corresponds approximately to  wavenumbers $k \simeq \ell/\eta_0 \simeq 0.2  \, \rm  h  Mpc^{-1}$ ($\eta_0 = \int_0^{a_r} dt/a$ is the conformal time at the current epoch).  The smallest scale that can be probed by low-redshift galaxy clustering data is also comparable  $k < 0.1 \, \rm h Mpc^{-1}$ as the density perturbations become non-linear at smaller scales. 

The angular power spectra for temperature, polarisation (E mode) and lensing potential along with the matter power spectra  are computed in CLASS codes for a given set of input parameters. The posteriors for these parameters are obtained using Montepython \cite{MontePy3_2018} implementation of MCMC. We choose Gaussian priors for the following  seven parameters:
\begin{table}[H]
    \centering
	\begin{tabular}{||c c c c c||}
	    \hline
	    Parameter & Mean & Std. Dev. & Lower lim & Upper lim\\
	    \hline
	    100 $\Omega_b h^2$ & 2.2377 & 0.015 & \_ & \_\\
	    $\Omega_{DM}$ & 0.255 & 0.0026 & \_ & \_\\
	    100 $\theta_s$ & 1.0411 & 0.0003 & \_ & \_\\
	    $n_s$ & 0.9659 & 0.0042 & \_ & \_\\
	    $\log{(10^{10} A_s)}$ & 3.0447 & 0.015 & \_ & \_\\
	    $\tau$ & 0.0543 & 0.008 & 0.004 & \_\\
	    $10^{-5} r$ & 12.0 & 100.0 & 1.0 & 100.0 \\
	    \hline
	\end{tabular}
	\caption{Parameter set and their priors}
\end{table}
MCMC optimisation with combined likelihoods of Planck temperature, polarisation (E mode), Lensing angular power spectra along with the BAO data is performed with $100$ chains of $100000$ steps each.  As we have no  prior information about the covariance of $r$ with the standard 6 parameters,  we use the $\Lambda$CDM covariance matrix with $r$ assumed to be independent for the first run. The ensuing chain is analysed to compute the covariance of $r$ with other parameters. The covariance matrix thus obtained is used for the subsequent runs. The average acceptance rate is $\simeq 0.19$ and the radius of convergence $R \simeq 0.1$. The best fit values and 1- and 2-$\sigma$ errors on the estimated parameters are shown in Table~\ref{tab:paramval}

\begin{figure}[H]
    \centering
    \includegraphics[width = 0.88\textwidth, trim = {0.5mm 1.5mm 0mm 0.5mm}, clip]{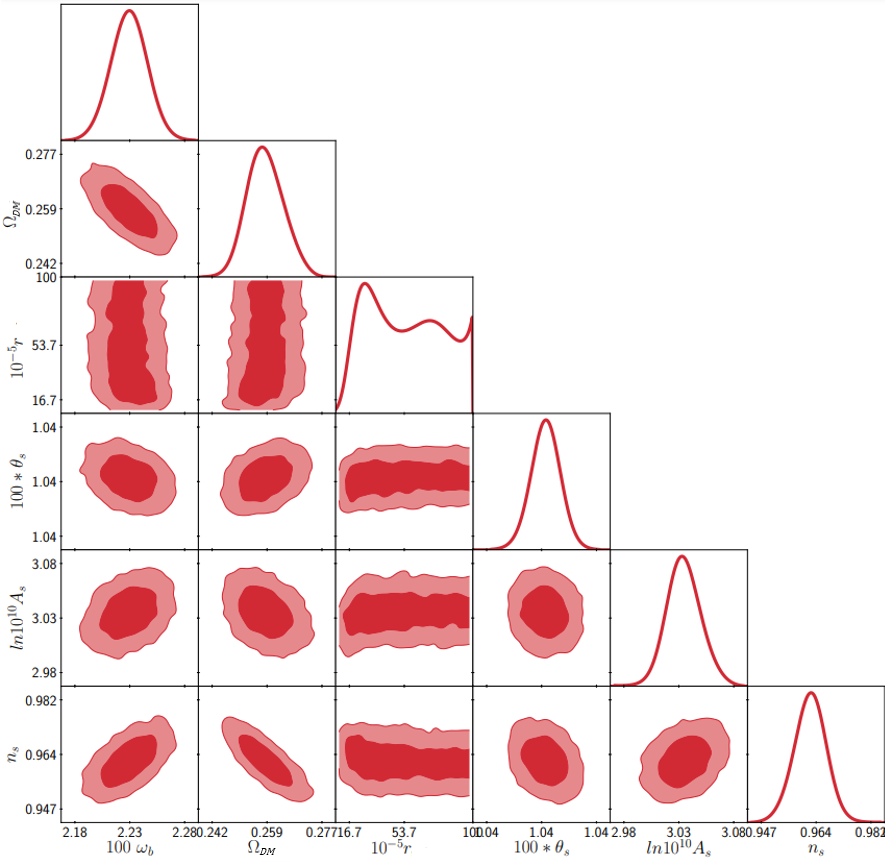}
    \caption{The Figure displays the contours and posterior probabilities of parameters from joint analysis of  CMB and BAO data sets.  1- and 2-$\sigma$ contours  correspond to deep and light red colours, respectively. The posterior probability of $r$ falls sharply for $r \lesssim 10^6$ but is flat for larger values of $r$, which shows the data  is consistent with $r \rightarrow \infty$ or the CDM model.}
    \label{fig:mcmc_result}
\end{figure}

\begin{table}[H]
    \begin{tabular}{|l|l|l|l|l|l|l|}
    \hline
    Parameters        & 2 $\sigma$ > & 1 $\sigma$ > & Best fit   & Mean       & 1 $\sigma$ < & 2 $\sigma$ < \\ \hline
    100 $\Omega_{b} h^2$ & 2.1989 & 2.2160 & 2.2233 & 2.2331 & 2.2505 & 2.2670   \\ \hline
    $\Omega_{DM}$   & 0.248 & 0.253 & 0.260 & 0.259 & 0.264 & 0.270   \\ \hline
    $10^{-5}r$   & -      & -      & 89.726 & 54.148 & -    & -         \\ \hline
    100 $\theta_{s}$ & 1.0419 & 1.0422 & 1.0424 & 1.0425 & 1.0427 & 1.0430   \\ \hline
    ln $10^{10}A_{s}$   & 3.0035 & 3.0188 & 3.0265 & 3.0338 & 3.0489 & 3.0649   \\ \hline
    $n_{s}$            & 0.9515 & 0.9567 & 0.9627 & 0.9623 & 0.9675 & 0.9731   \\ \hline
    $\tau_{reio}$    & 0.0388 & 0.0458 & 0.0512 & 0.0531 & 0.0602 & 0.0681   \\ \hline
    $H_0$                & 65.426 & 66.241 & 66.828 & 67.079 & 67.913 & 68.753   \\ \hline
    $\sigma_8$           & 0.7999 & 0.8065 & 0.8099 & 0.8127 & 0.8189 & 0.8256   \\ \hline
    \end{tabular}
    \centering
    \caption{Best fit values of the estimated parameters are displayed along with the 1- and 2-sigma errors computed from  their posterior probabilities. The posterior probability of  $r$ doesn't allow the determination of error on this parameter (Figure~\ref{fig:mcmc_result}).}
    \label{tab:paramval}
\end{table}

In Figure~\ref{fig:mcmc_result} we show the results of the MCMC analysis using the  combined  Planck 2018 CMB data along with the 
SDSS BAO data. The analysis yield an approximate lower bound: $r \gtrsim 10^6$. The lower limit on $r$ implies  the data is compatible with the $\Lambda$CDM model. For $r = 10^6$, the corresponding free-streaming scale $k_{\rm fs} \simeq 0.9$\, \rm h Mpc$^{-1}$ and 
the matter-radiation equality epoch, $a_{\rm eq} \simeq 2.9 \times 10^{-4}$ (Eq.~\eqref{eq:a_eq}). From the contour plots between
$r$ and other cosmological parameters, we note that the parameter $r$ is not constrained by the priors on other parameters.

\subsubsection{Collapsed HI fraction}
As noted above, the CMB and galaxy clustering data  probe scales  $k \lesssim 0.2 \, \rm h Mpc^{-1}$. The  WDM  models could deviate significantly from the $\Lambda$CDM model at small  scales.  Some cosmological probes such as Weak gravitational lensing and Lyman-$\alpha$ forest data allow probe of scales $k \lesssim 5 \, \rm Mpc^{-1}$ (e.g. \cite{mcquinn2016evolution,Bartelmann:1999yn} and references therein). However, comparing our results with  these data sets require extensive modelling 
which we consider beyond the scope of the current paper.

From absorption studies of high-redshift,   QSOs,  SDSS DR9 report the  detection of  nearly  7500 Damped Lyman-$\alpha$ absorbers. This allows one to 
estimate  precisely the average mass density of  neutral hydrogen (HI) in the redshift range $2 < z < 5$ (see \cite{SDSS_dr7,Peroux_HI} and reference therein; for more details see e.g. \cite{peebles:1993}).  The mass density  of HI can be related to the collapsed fraction of 
baryons and dark matter. This allows us to get an approximate measure of the 
minimum amount of collapsed fraction of the total matter in the redshift
range $2 < z < 5$. From the HI data one obtains, $\Omega_{\rm HI}(z) = \rho_{\rm HI}^{\rm coll}(z)/\rho_c(z)$, which gives the fraction of the collapsed neutral
hydrogen in terms of critical density of the universe $\rho_c$. The (minimum)  collapsed fraction is given by: $f_{\rm coll}(z) = \rho_c(z)/\rho_b(z) \Omega_{\rm HI}(z)$, where $\rho_b$ is the background energy density of  Baryons (for more details   
see section 5 of  \cite{Sarkar_2016}).

For computing the collapsed fraction, $f_{\rm coll}(M,z)$,  we integrate 
 the Sheth-Tormen mass function \cite{1999MNRAS.308..119S} above a certain mass threshold. It is not straightforward to compare the theoretical  collapsed fraction  with the damped
Lyman-$\alpha$ data because there is a large uncertainty in the masses of 
these clouds. The simulations suggest that these clouds could be  proto-galaxies with  baryonic  masses in the range $10^9\hbox{--}10^{10} \, \rm M_\odot$ (e.g. \cite{2008MNRAS.390.1349P}).
However, some recent observations suggest that the mass could be as high as 
$10^{12} \, \rm M_\odot$ at $z \simeq 2.5$. (\cite{2012JCAP...11..059F}).  For the present 
work, we assume two halo masses $10^{10}$ and $5\times 10^{10} \, \rm M_\odot$ as the threshold
masses for the formation of Damped Lyman-$\alpha$ clouds. We compute the collapsed fraction for comparison with the  data by 
integrating the mass function with  threshold mass  as lower limits. 

\begin{figure}[H]
    \centering
    \includegraphics[width = 0.49\textwidth]{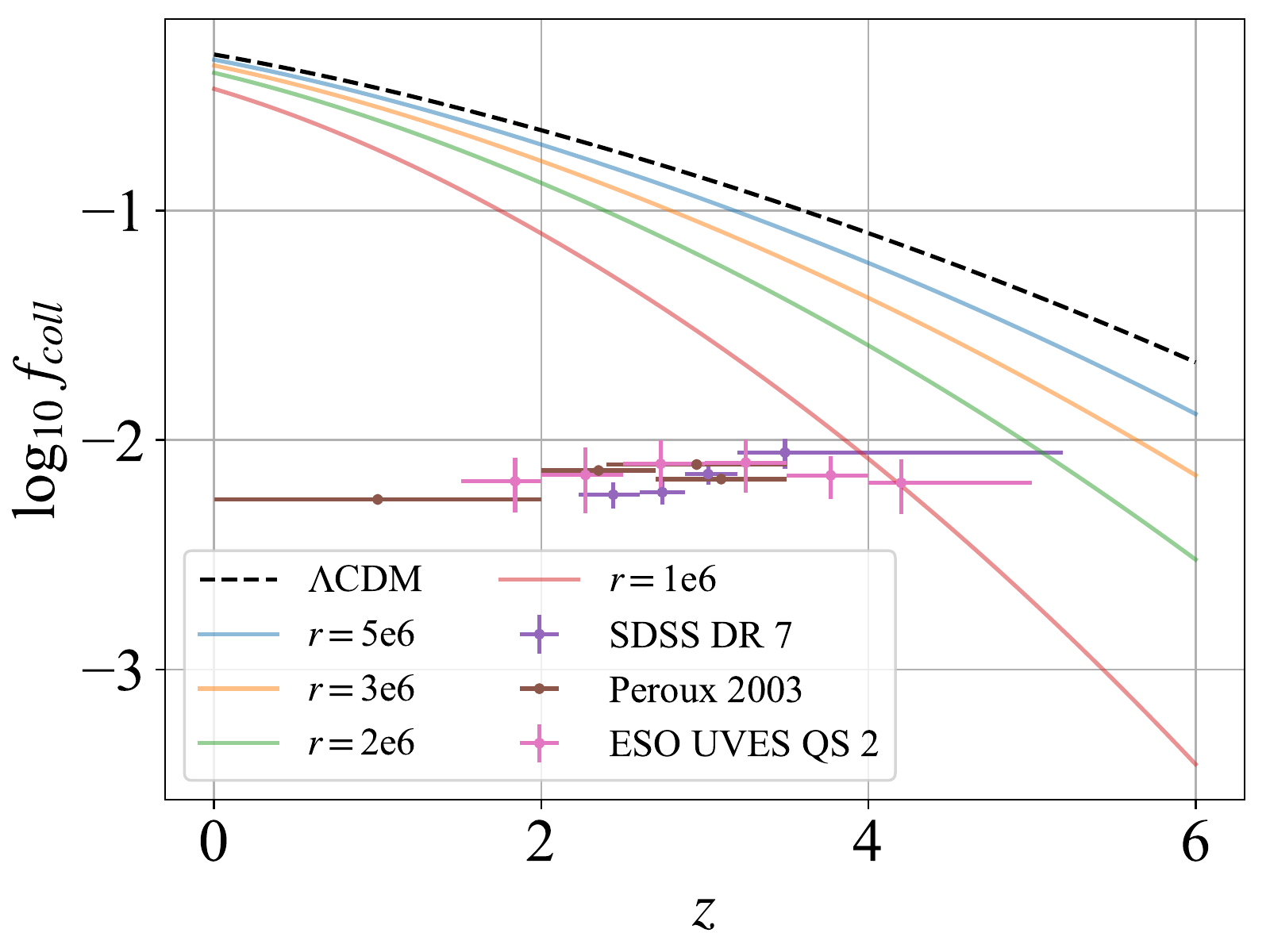}
    \includegraphics[width = 0.49\textwidth]{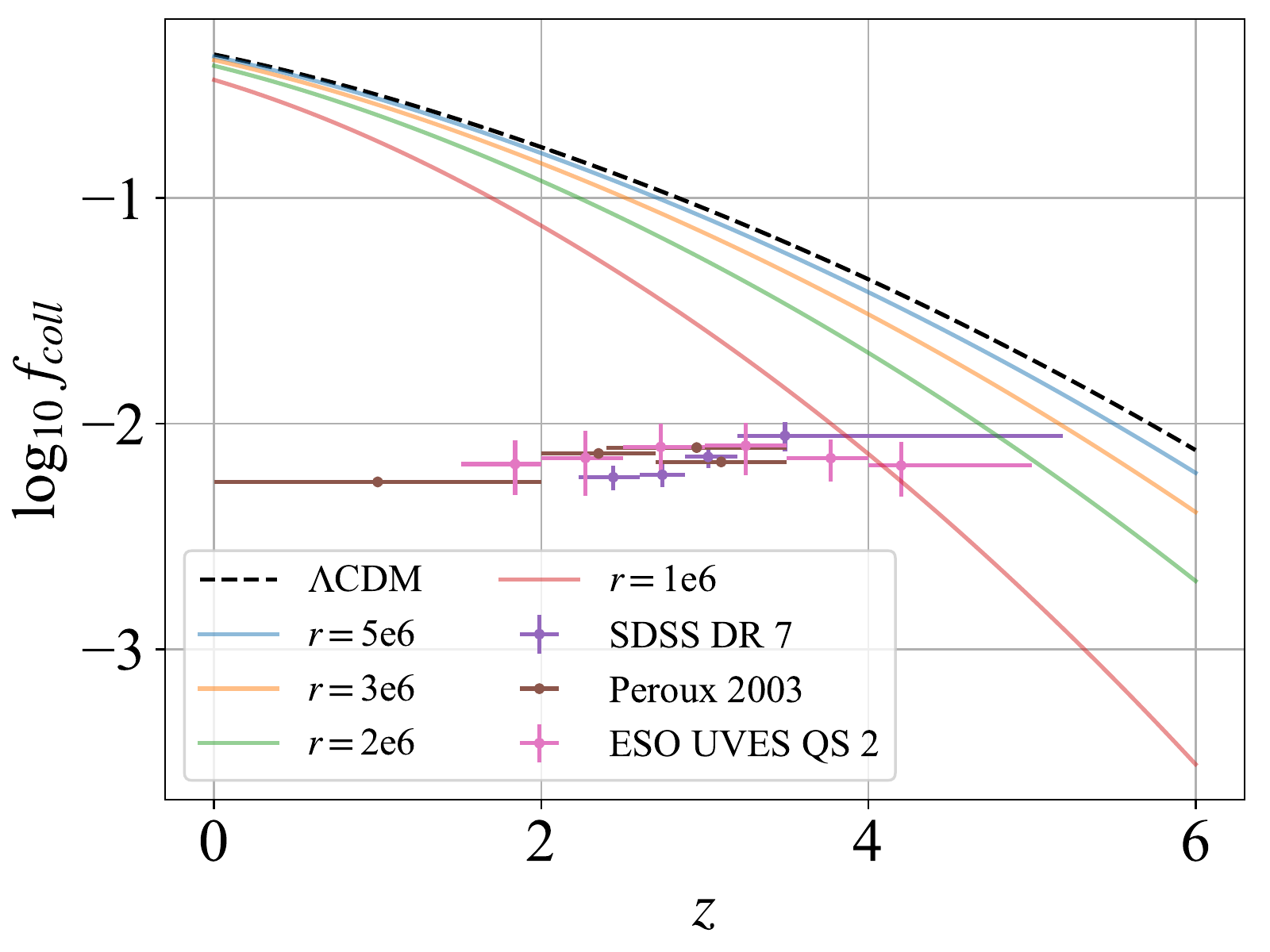}
    \caption{The two panels show the collapsed fraction as a function of redshift for different values of $r$. The two panels correspond to  the following mass thresholds: $M = 10^{10} \, \rm M_\odot$ (Left Panel) and $M = 5 \times 10^{10} \, \rm M_\odot$ (Right Panel). The  collapsed fraction from the HI data is also shown. Different curves correspond to the total collapsed fraction above a threshold mass. The HI data provides a lower limit to the collapsed fraction while comparing with theoretical models.}
    \label{fig:press_schechter_results}
\end{figure}

In Figure~\ref{fig:press_schechter_results} we display the collapsed fraction inferred from HI data against our models. The Figure shows that the evolution of collapsed fraction is an excellent diagnostic of small scale power,  as the collapsed fraction for models with lower $r$ shows significant decrement at high redshifts. As the HI data gives a lower limit to  the collapsed fraction, all the models that are well above the HI data could  be deemed to acceptable.    We can  see that the collapsed fraction for WDM model corresponding to $r = 10^6$ doesn't meet this requirement  at high redshifts. Therefore, this model   can be ruled out. Much better bounds can be obtained with higher redshift data and greater information on the mass range of damped Lyman-$\alpha$ clouds. The constraint  on $r$ is  of the same order as the one  obtained from Planck CMB and BAO datasets. 

\section{Summary and Conclusion}

\label{sec:concl}
\label{sec:conclusion}

In Figure~\ref{fig:param_space} we show the constraints  in $m_L$--$a_d$ space for $m_H = 100 \, \rm GeV$. The Figure displays  the entire parameter
space available in this case: $m_L \simeq 100\hbox{--}10^{11} \, \rm eV$ and $a_d \simeq 10^{-14}\hbox{--}10^{-4}$. The Figure
also shows curves corresponding to a constant  $\langle \sigma v \rangle $  compatible with the  Planck best-fit $\Omega_m$.
We also display the allowed region  arising from  the total radiation fraction of decay products to be less than 1\%.   A lower bound on $r$ also constrains this fraction and  we note that the data yields a stronger constraint. 
We summarize the main findings that follow from the figure. We also argue how
the parameter space shown in Figure~\ref{fig:param_space}  can be generalized to other values of $m_H$. 

\begin{itemize}
    \item [(a)]  The minimum  allowed WDM mass $m_L$  can be read off from  Figure~\ref{fig:param_space} for $m_H = 100 \, \rm GeV$. This  can be generalized to other WIMP masses. 
    The lower limit on $m_L$ corresponds   to $r = r_{\text{min}} \simeq 10^6$. We define $x_d = a_d m_H / T_0$. It follows from the definition of $r$ (Eq.~(\ref{eq:rdef})):

\begin{figure}[H]
    \centering
    \includegraphics[width = 0.85\textwidth]{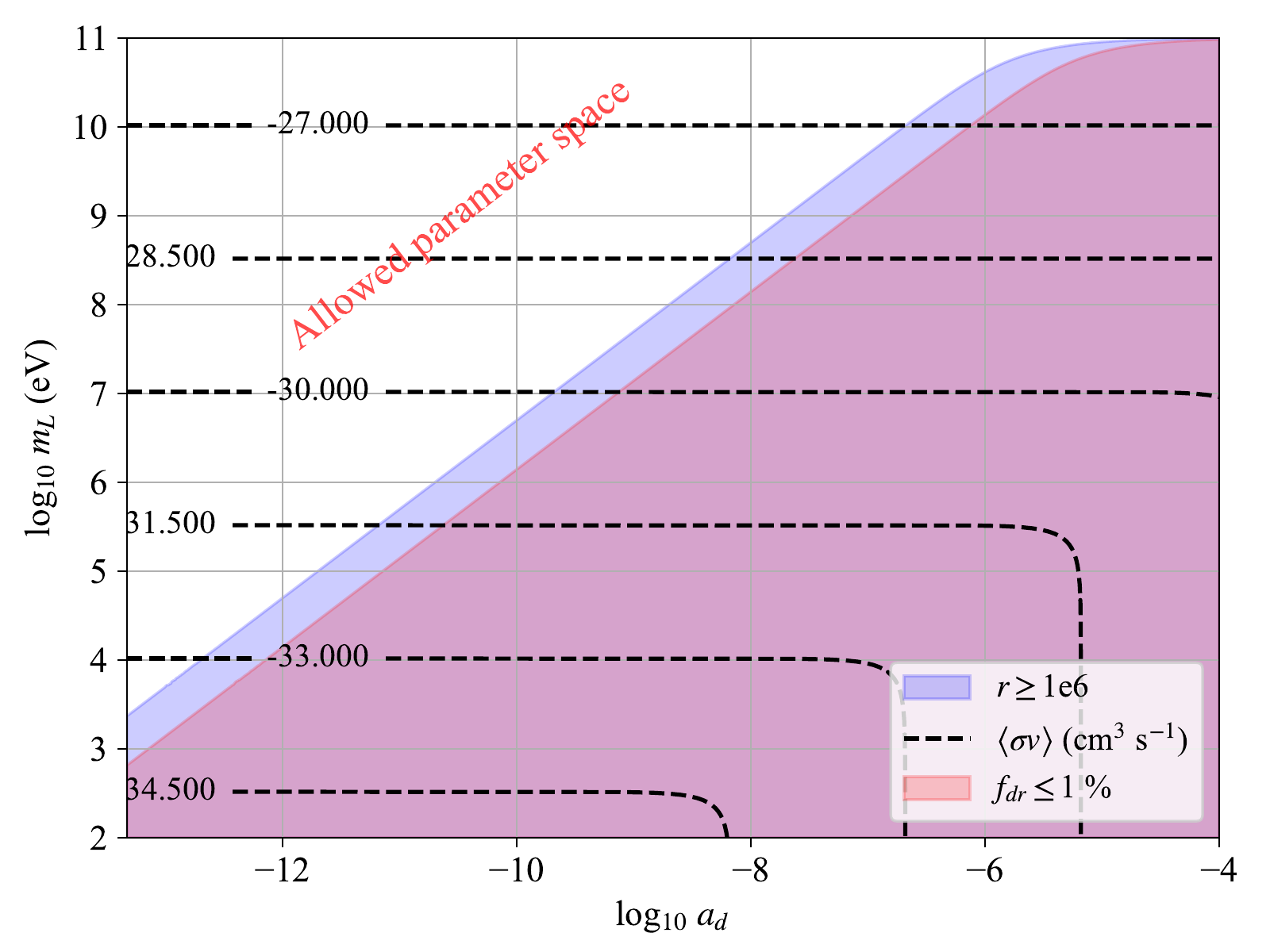}
    \caption{The figure shows the explored parameter space---$m_L$, $a_d$, and $\langle \sigma v \rangle$---and constraints
    from different cosmological observables for $m_H = 100 \, \rm GeV$. The blue  and the orange regions together are eliminated by  the constraint  $r< 10^6$ and it follows from the analysis of CMB, galaxy clustering, and HI data (see text for detail). The  orange  region alone  corresponds to the  parameter space eliminated by the bound $f_{\rm dr}<1\%$. 
    This region is shown for comparison and underlines our result that the constraints obtained from the  data are better than this bound.  The black dashed  lines display the Planck best-fit $\Omega_{\rm DM}$ for a given $\langle \sigma v \rangle$. The lines are labelled by values of  $\langle \sigma v \rangle$, e.g. -30 corresponds to $\langle \sigma v \rangle = 10^{-30} \, \rm cm^3 sec^{-1}$. These lines are obtained by   using Equation~(26) of \cite{Steigman_2012} and Eq.~\eqref{eq:const_2}. }
    \label{fig:param_space}
\end{figure}

\begin{align}
    m_{L, \text{min}} &= \frac{m_H}{a_d r_{\text{min}}} \Big( -1 + \sqrt{1 + (a_d r_{\text{min}})^2} \Big) \nonumber\\
    &\simeq \frac{m_H a_d r_{min}}{2} = \frac{T_0 x_d r_{min}}{2} \text{ for }a_d r_{min}<<1 \nonumber \\
    &= 2.35 \text{ keV } \Big( \frac{x_d}{20} \Big) \Big( \frac{r_{min}}{10^6} \Big)
    \label{eq:minmass}
\end{align}

Notice that $m_{L, \text{min}}$ is dependent on the product of $m_H$ and $a_d$.
As  Eq.~(\ref{eq:minmass}) shows,  the minimum mass   arises from the earliest possible decay of WIMP.
We assume the decay to be triggered after the freeze-out which implies a lower bound on $x_d$. We assume $x_d > a_* m_H / T_0 = x_* \simeq 20$ (e.g. Fig. 4 in \cite{Steigman_2012}). This prescription  ensures that    the decay
process   follows the freeze out epoch.
This assumption yields an approximate lower bound  $m_{L, \text{min}} \simeq 2.35$ keV.
We note that similar lower bounds on the mass of thermally-produced WDM have been obtained from the comparison of these models with small-scale cosmological data (e.g. Lyman-$\alpha$ and galactic  data)  at a range of scales  (e.g. \cite{WDM_mass_1,WDM_mass_2,viel2013warm,narayanan2000lyman}). 
The earliest possible decay of WIMP also yields  the maximum relic abundance $N$ from Eq.~\eqref{eq:const_2} (or minimum self annihilation  cross section $\langle\sigma v \rangle$ which can be obtained by computing  $N$ using Equation~26 of \cite{Steigman_2012}), the minimum comoving energy of the decay radiation, and the maximum fraction of the decay radiation:
\begin{align}
    N_{\text{max}} &= \frac{\Omega_{\rm DM} \rho_{c}}{m_{L, \text{min}}\sqrt{1 + (1/r_{\text{min}})^2}} \nonumber \\ &\simeq 3.99 \times 10^{-42} \text{ GeV}^3 (\hbar c)^{-3} \Big( \frac{20}{x_d} \Big) \Big( \frac{10^6}{r_{min}} \Big) \text{ for }r_{min} >> 1 \nonumber \\
    \langle \sigma v \rangle_{\text{min}} &= \left(2.4 \times 10^{-27} \text{ cm}^3 \text{ s}^{-1} \right) \frac{\rho_{c}}{m_H N_{\text{max}} h^2} \nonumber \\&= 4.87 \times 10^{-34} \text{ cm}^3 \text{ s}^{-1} \Big( \frac{x_d}{20} \Big) \Big( \frac{r_{min}}{10^6} \Big) \Big( \frac{100 \text{ GeV}}{m_H} \Big) \nonumber\\
    q_{\text{dr}, \text{ min}} &= \frac{m_{L, \text{min}}}{r_{\text{min}}} = 2.35 \text{ meV} \Big( \frac{x_d}{20} \Big) \nonumber \\
    f_{\text{dr, max}} &= \frac{\Omega_{\rm DM}}{(\Omega_{\gamma}+\Omega_{\nu})\sqrt{1+r_{\text{min}}^2}} = 0.28 \% \left( \frac{10^6}{r_{min}} \right)
    \label{eq:mincs}
\end{align}
We notice the maximum radiation fraction  $f_{\rm dr} \simeq 0.28 \%$. In addition, the WDM contributes  around 0.14\%  to the radiation density (see section \ref{sec:submreq} for detail). 
This is consistent with   the bound on  the fraction of radiation density  displayed in Figure~\ref{fig:param_space}. 
\item[(b)] Figure~\ref{fig:param_space} shows  the range of allowed $a_d$. The  minimum decay redshift  $1/a_d \simeq 10^6$ which arises from the requirement  that $m_L < m_H $ for $m_H = 100 \, \rm GeV$. The allowed range of $a_d$ follows from many related constraints: if $a_d$ is small and $m_L \ll m_H$, WDM is  highly relativistic at the time of decay.  This scenario provides the minimum lower bound as the WDM is hotter if the decay occurs later.  As $a_d$ is increased or the decay occurs at later times, the available parameter space shrinks as  the lighter WDM particle  are too hot  and therefore prevent growth of structures.   We note that the decay radiation could be photons if $z_d > 10^6$. In this case, the excess radiation gets thermalized with the plasma
and doesn't result in spectral distortion of the CMB spectrum (see e.g. \cite{2014PTEP.2014fB107T,chluba2019new,kogut2019cmb,peebles:1993}). However, the decay into photons at $z > 10^{10}$ can alter the light element abundance during primordial nucleosynthesis, which provides another constraint on such energy injection \cite{2017JCAP...03..043P}.
\item[(c)] In Figure~\ref{fig:param_space} we also show the contours of  velocity-weighted cross-section $\langle \sigma v \rangle$ that give the correct mass density for a given value of $m_L$ and $a_d$.  The intersection of these curves with the allowed region yields the set of parameters that satisfy all constraints. The allowed parameter space permits the range: $\langle \sigma v \rangle \simeq 10^{-26}\hbox{--}10^{-34} \, \rm cm^3 sec^{-1}$; Eq.~(\ref{eq:mincs}) gives the minimum value 
of $\langle \sigma v \rangle$ and its scaling with $m_H$.  Unlike the case of a stable WIMP which requires
the self-annihilation cross section to lie in a narrow range nearly independent of the mass of WIMP (e.g. \cite{Steigman_2012}), our proposed scenario expands the allowed parameter space by many orders of magnitude. We note that, given the range of self-annihilation cross-sections we explore in this 
paper, the particle interactions could be very different from a conventional WIMP. Even though such  particles could fall within the WIMP family of models,  it is conceivable such a particle 
arises from other physics.
\end{itemize}

 In our analysis, we assumed the WIMP to have zero  velocity at the time of decay which resulted in  WDM phase-space distribution function to be a delta function.  This is a good assumption as the decay particles are either highly relativistic at the time of decay or 
 the decay occur late enough (the WIMP velocity decays as $1/a$ after the kinetic decoupling) to justify this assumption. In either case,
 the WIMP velocity is negligible as compared to the speeds of the decay products and therefore has negligible impact on our results.  Our assumption  also renders the problem analytically tractable and less expensive to implement numerically (CLASS runtime for the WDM model $\simeq 0.5$ sec, which is comparable to the  CDM model). One possible extension of our model would be to start with the more realistic  Fermi  distribution for the WIMPs.

  In our analysis, we assume the decay of the particle to be instantaneous. In an expanding universe, instantaneous decay corresponds to the situation, $\tau \ll t$. It is possible to consider extended decay or $\tau \gtrsim t$ (e.g.  by extending the formalism discussed by \cite{1990PhRvD..41..354B} for photons). In our analysis, we consider the processes of freeze-out and decay separately. A more comprehensive  treatment would entail considering WIMPs as particles with a lifetime $\tau$  and solving the Boltzmann equations pertaining to their annihilation and decay simultaneously with the evolution of the background and the perturbed components of the multi-component fluid, which is  be beyond the scope of this paper. 

However, the results from the  instantaneous decay model allow us to qualitatively discern the outcome of models with larger $\tau$.  In the instantaneous decay model, all the WDM particles produced  have the same coldness parameter $r$ corresponding to $m_L$ and $a_d$. If  the lifetime $\tau$ pertains to a scale factor $a_d$, the WDM particles will have a distribution of coldness parameters and the change in the power spectrum will be determined by a weighted mean
of  coldness parameters over the time of decay $\tau$.  The power suppression at any scale is determined by the 
velocity of the decay product  at the time of horizon entry. The impact of finite decay time would be 
to spread the power  suppression over  a larger range of scales, which could lower the bound on $m_L$.

Many alternative models  to the WIMP-based CDM particle have been proposed, e.g. the WDM or ULA model (e.g. \cite{Marsh:2015xka,viel2013warm} and references therein).  However, even though our model is akin to non-thermal WDM scenario, it
belongs in the WIMP-inspired family of models. In particular, it permits parameter space in the mass range  $m_H < 100 \, \rm GeV$ which fell out of favour with improved 
sensitivity of XENON experiments \cite{PhysRevLett.121.111302,PhysRevLett.118.251302}. Our model  also predicts the presence
of a lighter particle whose presence can be revealed by  the cosmological data.  The current data puts a stringent lower bound
on the mass of this particle, $m_L > 2.35 \, \rm keV$.  Our proposed scenario  could also  alleviate  some small-scale issues  (e.g. cuspy profiles or missing satellites of the Milky way) of the stable WIMP model. 

In addition, as the model uses  parameter space 
of  SUSY-based models, the heavier particle  and/or its decay products  might be detectable in collider experiments.

The past few decades have seen tremendous improvement in cosmological data  and experimental sensitivity of XENON-based dark matter
experiments. While cosmological data has thrown light on the nature of dark matter, its detection has eluded us. This has motivated 
theorists to move beyond the most-favoured 
model based on WIMPs. Our work suggests an alternative model that can be accommodated within the WIMP paradigm.

\acknowledgments

One of us (AP) sincerely thanks Dr. Harvinder Kaur Jassal for her invaluable advice and assistance during the research process. Additionally, we would like to extend our acknowledgement to Dr. Jasjeet Singh Bagla for his insights and suggestions regarding the halo mass function analysis. 

\appendix
\section{Appendix: CLASS implementation}

The default configuration for both our own Python code and CLASS implementation (in explanatory.ini, the input parameter file): (i) Linear scalar perturbations in Newtonian gauge with adiabatic initial conditions (ii) Evolution from scale factor $a = 10^{-14}$ to $a = 1$ (iii) Primordial helium fraction $\rm YHe = 0.24$ instead of default 'BBN' for consistency of CLASS run with own Python code (iv) Power law primordial power spectrum with index $n_s = 0.9660499$ for Gaussian fluctuations (v) 5 species: dark energy, dark matter, photons, baryons and massless neutrinos (vi) Spatially flat model ($\Omega_{k} = 0$) with the dark energy  modelled as cosmological constant that only affects the background  (vii) $\Omega_b = 0.045$ (viii) $\Omega_{\rm cdm}$ is set to $0$, $\Omega_{\rm wdm}$ to $0.255$ with  new variables $a_{d}$ and  $r$  (ix) No massive neutrinos are considered and the only ultra relativistic species are the three  massless neutrinos  with $\Omega_{\nu} = 3.56793 \times 10^{-5}$. The other settings are the same as default settings for CLASS (which is configured for $\Lambda$CDM model). 

Here we list the CLASS files and the corresponding changes made in them to implement our  model. We  added two more species---warm dark matter and decay radiation---in the input, background and perturbations files.
\subsection*{input.c}
\begin{itemize}
    \item In {input\_read\_parameters\_species()}, we defined local variables {flag4, flag5, param4, param5, has\_wdm\_userdefined} in lines 2301-2313. The values for the parameters $\Omega_{wdm}$, $r_{wdm}$, $a_{d, wdm}$ were read from the input in lines 2469-2508 with appropriate checks and stored in the structure {background *pba}. The value for $\Omega_{\text{dr}}$ was computed using \eqref{eq:0_ord_quant_rad} and stored.
    
    \item The contribution of $\Omega_{wdm}$ and $\Omega_{\text{dr}}$ to the total energy budget $\Omega_{\text{total}}$ was added in lines 3207-3208.

    \item Default values for $\Omega_{cdm}$, $\Omega_{wdm}$, $r_{wdm}$, $a_{d}$ and $\Omega_{\text{dr}}$ were set to zero in lines 5738-5744.
\end{itemize}

\subsection*{background.h}
\begin{itemize}
    \item The variables $\Omega_{wdm}$, $r_{wdm}$, $a_{d}$ and $\Omega_{\text{dr}}$ were added to the structure background in lines 72-75.
    \item Memory was allocated for index variables of  the background density and pressure  arrays for  the dark matter and decay radiation  in lines 174-177.
    \item Conditional variable has\_wdm was added in line 294. If its value is \_TRUE\_, it would mean both the WDM and decay radiation are present.
\end{itemize}

\subsection*{background.c}
\begin{itemize}
    \item In background\_functions(), local variables for density and pressure of WDM ($\rho_{wdm}$, $P_{wdm}$) and decay radiation ($\rho_{\text{dr}}$, $P_\text{dr}$) were added at lines 393-394.
    
    \item $\rho_{wdm}$, $P_{wdm}$, $\rho_{\text{dr}}$ and $P_{\text{dr}}$ were computed using Eqs.\eqref{eq:0_ord_quant_CDM}, \eqref{eq:0_ord_quant_WDM} and~\eqref{eq:0_ord_quant_rad} in lines 447-460 based upon the condition $a<a_d$. Prior to  the decay, $\rho_{\text{dr}}$ and $p_{\text{dr}}$ were set to zero. The computed values were then stored in corresponding time arrays in structure background in lines 461-464. Their contribution to total density: $\rho_{wdm} + \rho_{\text{dr}}$, total pressure: $P_{wdm} + P_{\text{dr}}$, matter density: $\rho_{wdm} - 3P_{wdm}$ and radiation density: $3P_{wdm} + \rho_{\text{dr}}$ were added in lines 466-469.
    
    \item In background\_w\_fld(), $\Omega_{wdm}$ and $\Omega_{\text{dr}}$ are added in $\Omega_m$ and $\Omega_r$ in lines 724-725. Matter-radiation equality is calculated as $a_{eq} = \Omega_r/\Omega_m$. A more accurate calculation uses Eq.~\eqref{eq:a_eq}. We can check that Eq.~\eqref{eq:a_eq} reduces to $a_{\rm eq} = \Omega_r/\Omega_m$ for $r_{wdm} \rightarrow \infty$.
    
    \item In background\_indices(), default value of has\_wdm is set to \_FALSE\_ in 1004 and is assigned \_TRUE\_ only if $\Omega_{wdm}$ is non-zero in 1020-1021. The background density and pressure array indices are initialised in lines 1079-1083.

    \item In background\_solve(), non free-streaming dark matter fraction does not receive any contribution from WDM, see lines 2150-2162.

    \item In background\_initial\_conditions(), the initial $\rho_{\text{r}}$ receives contribution from neither the dark matter, which is cold prior to the decay, nor the decay radiation (see lines 2254-2270).

    \item In background\_output\_titles(), column titles 'rho\_wdm' and 'rho\_decay\_rad' are added in lines 2502-2503.

    \item In background\_output\_data(), the background time arrays for WDM and decay radiation are written, in lines 2579-2582.

    \item In background\_derivs(), WDM contributes $\rho_{wdm} - 3P_{wdm}$ to $\rho_M$, see lines 2700-2702.

    \item background\_output\_budget(), $\Omega_{wdm}$ and $\Omega_{\text{dr}}$ are printed out, see lines 2873-2878.
\end{itemize}

\subsection*{perturbations.h}
\begin{itemize}
    \item In structure perturbations, we added boolean variables for qualifying the presence of source $\delta$ and $\theta$ for WDM and decay radiation in lines 240, 241, 255, 256. Index variables for the same were added in lines 291, 292, 311, 312.

    \item In structure perturbations\_vector, time array indices for $\delta, \theta, \sigma$ and maximum number  of moments for WDM and decay radiation were added in lines 477-484.
\end{itemize}

\subsection*{perturbations.c}
\begin{itemize}
    \item In background\_output\_data(), time array for over-density of WDM and decay radiation were written in lines 473-474 and 525-526.

    \item Column titles of $\delta$, $\theta$ and transfer function arrays of perturbations structure were stored in lines 565-566, 596-597 and 619-620.

    \item Boolean variables for qualifying the  presence of source $\delta$ and  $\theta$ and index variables of perturbations structure were initialised in perturbations\_indices().

    \item WDM and decay radiation were added in the computation of maximum moment for any species in lines 2808-2809.

    \item Column titles of $\delta, \theta, \sigma$ arrays of perturbations\_vector structure were stored in lines 3457-3462.

    \item Time array indices for $\delta, \theta, \sigma$ of perturbations\_vector structure were initialised in 4055-4060.

    \item Since we do not use any approximation schemes (tight coupling, ultra-relativistic, etc) for WDM or decay radiation, the values of $\delta$, $\theta$, $\sigma$  are recopied into the original array perturbations\_vector structure in lines 4517-4536.

    \item In perturbations\_initial\_conditions(), we define local variables frac\_wdm, $r$ and $a_d$. Radiation and matter density contributions from WDM and decay radiation are added in lines 5493-5497. frac\_wdm is initialised in line 5547, which is later used in the computation of $\alpha$---a conversion factor between synchronous and Newtonian initial conditions.

    \item The initial conditions for perturbations of both WDM and decay radiation are taken from  CDM initial conditions (Eq.~\eqref{eq:init_CDM}) in lines 5946-5957. 

    \item In perturbations\_einstein() line 6708, the metric perturbation $\Phi$ was computed using the 2nd perturbed Einstein equation which uses total $\rho\delta$ (Equation~(23b) in \cite{Ma_bert}):
    \begin{equation*}
        k\big(-\dot{\Phi}+\frac{\dot{a}}{a}\Psi\big) = 4\pi G a^2 \Sigma (\bar{\rho} + \bar{P})\theta
    \end{equation*}
    Instead, we use the 1st perturbed Einstein equation which uses total $(\rho + P)\theta$ (Equation~(23a) in \cite{Ma_bert}) to maintain consistency with our own codes.
    \begin{equation*}
        k^2\Phi + 3\big(\frac{\dot{a}}{a}\big) (\dot{\Phi} - \frac{\dot{a}}{a}\Psi) = 4\pi G a^2 \Sigma \bar{\rho} \delta
    \end{equation*}
    
    \item In perturbations\_total\_stress\_energy(), we add the WDM and decay radiation contributions to $\rho+P|_{\text{total}}, \rho \delta, P\delta (\rho+P)\theta$ and $(\rho+P)\sigma$ in lines 7097-7103. The WDM contribution to $\rho_m, \rho\delta|_m, (\rho+P)|_m$ and $(\rho+P)\theta|_m$ is added in lines 7105-7111. 

    \item In perturbations\_print\_variables(), local variables for $\delta$, $\theta$, $\sigma$ are defined and initialised with the computed time arrays of perturbations\_vector structure in lines 8451-8457, to be stored in lines 8678-8683, without any gauge transformation required since we use Boltzmann equations in Newtonian gauge. CLASS by default is implemented in synchronous gauge and transformed to the  Newtonian gauge, if required by input. Our WDM implementation will work only for Newtonian gauge input.
    
    \item In perturbations\_derivs(), we define and initialise local variables $r$ and $a_d$ with their corresponding values in structure background. For WDM, we compute the derivatives of $\delta, \theta, \sigma$ using Eq.~\eqref{eq:Boltz_CDM} for $a<a_d$ and Eq.~\eqref{eq:Boltz_WDM} otherwise. For decay radiation, we compute the derivatives of $\delta, \theta, \sigma$ using Eq.~\eqref{eq:Boltz_CDM} for $a<a_d$ and Eq.~\eqref{eq:Boltz_rad} otherwise.  
\end{itemize}

\section{Appendix: Freeze-out}
\label{app:fout}

The  fitting function given in Equation~26 of \cite{Steigman_2012}  was obtained for cross-sections in the range $10^{-27} \, \rm {\rm cm^{3} \, sec^{-1}} \lesssim \langle \sigma v \rangle \lesssim 10^{-15} \, \rm {\rm cm^{3} \, sec^{-1}}$. In our work, we use cross-sections in the range $10^{-34} \, \rm {\rm cm^{3} \, sec^{-1}} \lesssim \langle \sigma v \rangle \lesssim 10^{-26} \, \rm {\rm cm^{3} \, sec^{-1}}$. For smaller cross-sections the freeze-out occurs when the particle is still semi-relativistic. To address this issue
we numerically solve the relevant Boltzman equation to obtain the relation between 
the relic abundance and the thermally-averaged cross-section. We obtain the following fit:
\begin{equation}
    \label{eq:N_sigma_fit} \log_{10} \left( \frac{N}{1 \text{ GeV}^3 (\hbar c)^{-3}} \right) = -73.38 -0.95 \log_{10} \left( \left( \frac{m_H}{100 \text{ GeV}} \right) \left( \frac{\langle \sigma v \rangle}{1 \text{ cm}^3 s^{-1}} \right)\right)
\end{equation}


Eqs.~\eqref{eq:N_sigma_fit}  and~\eqref{eq:mincs} are   in good  agreement in the range $\langle \sigma v \rangle \in [10^{-27}, 10^{-34}] \rm cm^3 s^{-1}$. Using $\Gamma = n \langle \sigma v \rangle = H$, the  epoch of freeze-out  for $\langle \sigma v \rangle \simeq 10^{-26} \rm cm^3 s^{-1}$ is $x_* \simeq 25$  and for $\langle \sigma v \rangle \simeq 10^{-34} \rm cm^3 s^{-1}$, $x_* \simeq 5$. Therefore, for lower $\langle \sigma v \rangle \simeq 10^{-34} \rm cm^3 s^{-1}$, $x_* \sim 5$, we have semi-relativistic freeze-out, which does introduce some difference between  $N \propto \langle \sigma v \rangle^{-1}$ and \eqref{eq:N_sigma_fit}, which gives  $N \propto \langle \sigma v \rangle^{-0.95}$. However the difference is negligible. 
Eq.~\eqref{eq:mincs} provides a good fit to both  Equation~26 of \cite{Steigman_2012} and 
Eq.~\eqref{eq:N_sigma_fit}.



\bibliographystyle{JHEP}
\bibliography{lib}







\end{document}